\documentclass[aps,pre,twocolumn,superscriptaddress,showpacs,showkeys ]{revtex4-1}

\usepackage{graphicx}                   
\usepackage{hyperref}                   
\usepackage{amsmath,amssymb}            
\usepackage{epstopdf}
\usepackage{subcaption}					

\usepackage{color}
\definecolor{orange}{rgb}{1,0.5,0}
\definecolor{brown}{rgb}{0.65, 0.16, 0.16}
\definecolor{phlox}{rgb}{0.87, 0.0, 1.0}

\begin{document}
\author{M. N. Najafi}
\affiliation{Department of Physics, University of Mohaghegh Ardabili, P.O. Box 179, Ardabil, Iran}
\email{morteza.nattagh@gmail.com}
\title{Invasion percolation in short-range and long-range disorder background}
\author{S. Tizdast}
\affiliation{Department of Physics, University of Mohaghegh Ardabili, P.O. Box 179, Ardabil, Iran}
\author{J. Cheraghalizadeh}
\affiliation{Department of Physics, University of Mohaghegh Ardabili, P.O. Box 179, Ardabil, Iran}
\author{H. Dashti N.}
\affiliation{School of Physics, Korea Institute for Advanced Study, Seoul 02455, Korea.}
\begin{abstract}
In this paper we investigate the invasion percolation (IP) in imperfect support in which the configuration of imperfections is considered to be correlated. Three lattice models were engaged to realize this pattern: site percolation, Ising model and random coulomb potential (RCP). The first two models are short range interaction (SRI), whereas the last one includes coulomb like interactions which is pretty long range (long-range interactions, LRI). By examining various dynamical observables we show that the critical exponents of SRI IP are robust against the control parameters (temperature in the Ising model and occupation probability in site percolation), whereas its properties in the LRI (RCP) supports are completely different from the normal IP (i.e. on the regular lattice). Especially the fractal dimension of the external frontier of the largest hole converges to $1.099\pm 0.008$ for RCP IP, whereas it is nearly $\frac{4}{3}$ for SRI IP being compatible with normal IP. Additionally a novel dynamical crossover is seen in the RCP IP according to which the time dependence of all of the observables is divided to three parts: the power-law (small times), the logarithmic (mid time), and the linear (long time) regimes. The second crossover time is shown to go to infinity in the thermodynamic limit, whereas the first crossover time is nearly unchanged, signaling the dominance of the logarithmic regime. The observables become nearly constant in the thermodynamic limit for the long time, showing that it is a stationary phase.
\end{abstract}
\maketitle

\section{Introduction}
Invasion percolation (IP)~\cite{chandler1982capillary,wilkinson1983invasion} was first introduced in 1983 by Wilkinson and Willemsen to describe the slow displacement of one fluid by another in a porous medium~\cite{wilkinson1983invasion}. IP realizes the invasion phenomenon during which one phase invades the other one in a porous medium. Invasion percolation is divided into two general categories: trapping invasion percolation and non-trapping invasion percolation in which the defender fluid is incompressible and compressible respectively~\cite{stark1991invasion}. In a lattice setup this model is simple to define: one fills the system by the defender fluid and then the invader fluid (displacement) is injected into the environment. The stochastisity of fluid movement (due to stochastic properties of the voids in the porous media) is realized by an uncorrelated random generated variable. The main difference between IP and ordinary percolation is that it automatically organizes itself in a critical point, for which one uses the term self-organized criticality~\cite{de1978lois,lenormand1980description}. Many aspects and properties of IP in known in the community. For most important example is the well-known fact that the fractal structure of IP clusters (the cluster that is formed by invaded sites) is just similar to the one for the standard (site) percolation, e.g. the fractal dimension of the external frontier of largest hole is $D_f=\frac{4}{3}$~\cite{lenormand1985invasion,englman1986fragmentation,parkhurst1986better}. The other examples are three-dimensional IP~\cite{xu2008dynamics}, IP in correlated porous media~\cite{vidales1996invasion,babadagli2000invasion}, fractal growth dependence on the coordination number~\cite{knackstedt2002nonuniversality}, and many applications in the reservoir engineering~\cite{peter2018percolation}. Recently it was shown that, apart from the dynamical power-law behaviors, IP shows a dynamical crossover during which the autocorrelations change sign~\cite{tizdast2020dynamical}. For a good review on the theoretical and experimental development of IP see~\cite{feder1988fractals}. Despite of the intense research on the various variants of IP, and exploring its features and the properties, the effect of the background disorder has not been understood well yet. By the ``background disorder'' we mean the disorder in the configuration of the imperfections in the porous media. As a well-known fact, the porous media is formed by the sedimentation process during which some parts become permeable to the fluid, and some other impermeable, which we call the permeability field (PF). In fact the state of the porous media is partially described by PF, and knowing the pattern of it helps much in any prediction of the model that is employed to simulate the fluid propagation. It is not hard to convince one that the configuration of PF is not totally uncorrelated, since the sedimentation process can generally be correlated, i.e. as a dynamical growth process the correlation length has a chance to be non-zero. The effect of the correlated configuration of PF has not been convincingly understood yet in the literature and very limited attention have been paid to this issue, like IP in fractional Gaussian noise~\cite{mukhopadhyay2000calculation,knackstedt2001invasion,vidales1996invasion,babadagli2000invasion}. \\

The correlations that are created using the fractional Gaussian noise seems to be much artificial to be applicable for realistic situations. It is the aim of this paper to consider a wider range of models to capture the correlations in the support for two categories: short-range interaction (SRI) models and long-range interaction (LRI) one. This is done in a comprehensive and self-consistent method. To this end we consider the Ising and percolation models as well as the  RCP are employed to simulate the correlations between imperfections in the lattice. The interactions in the first two models are of short range nature (SRI), whereas the third model is pretty long-range (LRI). When the correlated lattices are constructed, we run the IP dynamics on top of them, avoiding the fluid to enter the impermeable regions (sites in our model). Our studies show that for the LRI model, the properties of the invasion cluster change drastically, whereas for the SRI models, the properties of IP does not considerably change.

The paper has been organized as follows: In the next section, the simulation performed is described. In the third section, we will describe and present numerical details and simulation results. We will close the paper with a conclusion. The paper includes an appendix.

\section{General Setup of the Problem}

\subsection{The IP model on the imperfect system}
Our model is defined on an $L \times L$ square lattice with permeable and impermeable sites. The pattern of impermeable sites (imperfections) are determined by three models in this paper: percolation, Ising and  RCP models to be described in the following. For the percolation and Ising models (SRI supports) the sites can have two possible states, represented by $s_i=+1$ or $-1$ showing that the site $i$ is permeable or impermeable respectively, whereas for the RCP $s_i=+1$ for all sites. The fluid can only pass through permeable sites, i.e. the clusters comprised by sites with $s_i=+1$. For the SRI supports (Ising- and percolation-correlated lattices) the configuration of $\left\lbrace s_i\right\rbrace_{i=1}^{L^2}$ is fixed using the Ising and site percolation models respectively, whereas for the LRI case the RCP realizes the size/quality of the pores and all sites are accessible. We describe these models in the following subsection. IP is run over the largest percolating connected cluster (PCC, comprised by $s=+1$ sites) with total $N$ sites, which is a cluster that connects two opposite boundaries of the lattice (for the  RCP case it is actually the original lattice). Once a PCC is extracted, the IP growth model is defined on top of it, defined as follows: $N$ uncorrelated random numbers $r$ in the range $[0,1]$ are distributed over the PCC, so that the state of the porous media is identified by $\left\lbrace r_i\right\rbrace_{i=1}^N$. The dynamics starts from a middle point of the lattice $i_0$ (if it does not belong to the percolating cluster, we move in a random direction and consider the first site belonging to the PCC as the starting point) where the fluid is injected. At the next step, the invader moves to a neighbor of the injected point (say the site $j$) with smallest $r$, i.e. $r_j=\text{min}\left\lbrace r_i\right\rbrace_{i\in \partial S(1)} $, where $S(m)$ is the set of the infected (filled by the invader) sites up to the step $m$, and $\partial S$ is the set of neighbors of the infected sites $S$. If two or more neighbors are equal in the parameter $r$, one of these minimums are randomly chosen. In the step $m+1$ the fluid enters the site $j\in \partial S(m)$, which is identified by the condition $r_j=\text{min}\left\lbrace r_i\right\rbrace_{i\in \partial S(m)} $ (if there are more, it is selected randomly between the set of neighboring sites with minimum $r$). The ``time'' is defined as the integer part of $\frac{m}{10}$. The process goes ahead until two opposite boundaries are touched by $S(m_{\text{max}})$. In the ordinary IP a phase transition occurs at this point to a phase where the invader fills the space, where IP shows power-law behavior~\cite{tizdast2020dynamical}. This IP has the same properties as the ordinary site percolation at the critical point. IP shows also power behavior with respect to time for various observables, and there are some scaling relations between the quantities.\\

The quantities that we analyze here are: \\
** The loop length ($l$) is defined as the length of the loop surrounding the cluster, so that $l(t)\equiv \sum_{j=1}^N\delta_{j,\partial S(t)}$, where $\delta_{j,\partial S}=1$ when $j\in S$, and zero otherwise, i.e. it is the number of sites on the boundary of $S(t)$.\\ 
** The loop and mass gyration radius ($r_{l}(t)$ and $r_m(t)$ respectively) are defined by
\begin{equation}
\begin{split}
r_l(t)^2 = \frac{1}{l(t)}\sum_{i \in \partial S(t)} {\left[ {{{({x_i} - \bar x_{l})}^2} + {{({y_i} - \bar y_{l})}^2}} \right]}, \\
r_m(t)^2 = \frac{1}{S(t)}\sum_{i \in S(t)} {\left[ {{{({x_i} - \bar x_{m})}^2} + {{({y_i} - \bar y_{m})}^2}} \right]}, 
\end{split}
\end{equation}
where ${x_i}$ and $ {y_i} $ are the Cartesian coordinates of the site $i$ and $ (\bar x_{l},\bar y_{l}) $ is the center of mass for loop gyration radius and  $ (\bar x_{m},\bar y_{m}) $ is the center of mass for mass gyration radius.\\
** The roughness ($w$) is defined by
\begin{equation}
w(t)^2 = \frac{1}{l(t)}\sum_{i \in \partial S(t)} (r_i - \bar r)^2 ,
\end{equation}
that ${r_i} \equiv (x_i^2 + y_i^2)^{\frac{1}{2}}$ and $\bar r \equiv \frac{1}{l(t)}\sum_{i \in \partial S(t)}  (x_i^2 + y_i^2)^{\frac{1}{2}}$.\\

\begin{figure}
	\centerline{\includegraphics[scale=.33]{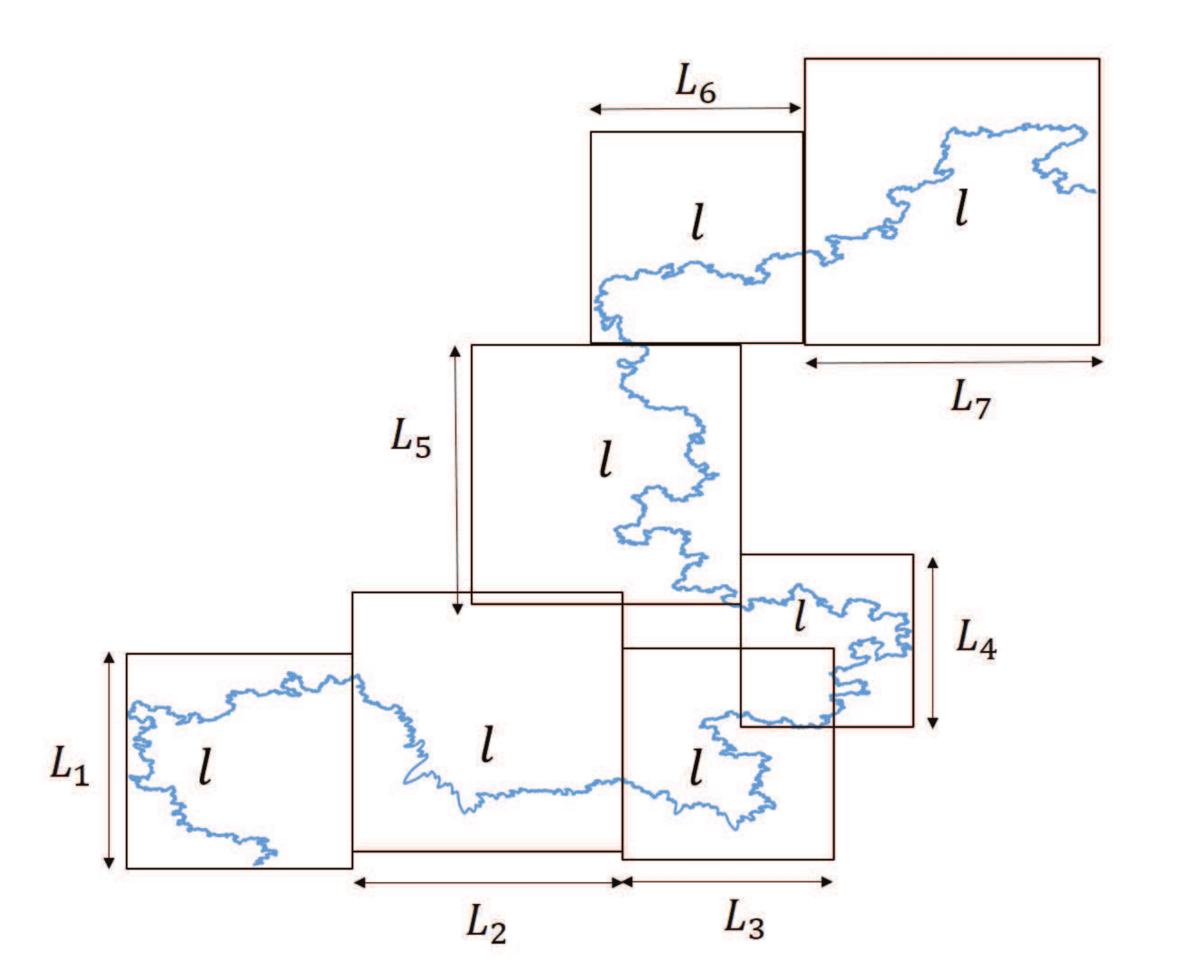}}
	\caption{The procedure of calculating the fractal dimensional(box counting).}
	\label{fig:bd}
\end{figure}

In the standard IP model $x$, $x = w,l,{r_l},{r_m}$ shows power-law behavior with time, reflected in the following relation\\
\begin{equation}
	\left\langle x \right\rangle  \propto { t  ^{{\alpha _x}}}
\end{equation}
where $\alpha_r$ accounts for the type of diffusion, i.e. for $\alpha_r<\frac{1}{2}$ ($>\frac{1}{2}$) we are sub (super) diffusion regime, whereas for $\alpha_r=\frac{1}{2}$ we are right in the normal diffusion regime. Two types of fractal dimension can be defined: dynamic fractal dimension (DFD, shown by $D^D_f$), and static fractal dimension (SFD, shown by $D^S_f$). The former is defined via the dynamical relation between $l$ and $r$, i.e. $\left\langle \log l(t)\right\rangle=D^D_f\left\langle\log r_l(t) \right\rangle +cnt$ ($\left\langle ... \right\rangle $ being the ensemble average). For the latter case (SFD) we consider the largest hole of the system in the percolation time ($t_{\text{perc}}$), and extract the fractal dimension of the boundary of the largest hole using the box-counting (BC) scheme to find $D^S_f$, shown schematically in Fig.\ref{fig:bd}. The holes of the system are obtained using the Hoshen Kopelman (HK) algorithm~\cite{hoshen1976percolation}. Importantly, using HK algorithm we extracted and analyzed the largest hole for which the fractal dimension is obtained to be $D^S_f(\text{ordinary IP})=1.33\pm 0.01$ as expected~\cite{tizdast2020dynamical}.\\ 

In the next subsection we explain the models that simulate the correlated pattern of the imperfections.

\subsection{Supports: Percolation, Ising and RCP correlated lattices}
Let us give a brief description of the models which are used to simulate the correlated lattices served as the support of IP. The models are listed bellow:\\

\textbf{\textit{1-Site Percolation}}, which is one of the most important, and at the same time simply defined examples in the out-of-equilibrium critical phenomena, which here is employed to fix the impermeable site configuration $\left\lbrace s_i\right\rbrace_{i=1}^{N}$. In this model we set this configuration at random with an external parameter $0\le p\le 1$: for each site $i$ we set $s_i=+1$ with probability $p$, and $s_i=-1$ with probability $1-p$. Many properties of this model are known thanks to probability theory~\cite{werner2007lectures,smirnov2001critical, langlands1994conformal, cardy2001conformal}, $c=0$ conformal filed theory~\cite{mathieu2007percolation}, coulomb gas~\cite{clarke2008impact}, and $Q$-state Potts model~\cite{potts1952some, wu1982potts, essam1979potts}. In this model, the for $p$ values lower than a critical threshold $p_{c}$, there is no percolated cluster, whereas for the case $p\ge p_c$ with almost definitely we have percolation cluster. The critical as well as off-critical properties of this model have been investigated widely in the literature~\cite{werner2007lectures,smirnov2001critical}. For the square lattice it is known that $p_c\sim 0.5927$~\cite{malarz2005square, majewski2006square}. In our paper we consider only the percolated clusters as a host for the IP dynamics. The dynamical aspects of some statistical models have been already studied on the site percolation systems, like the sandpiles~\cite{najafi2016bak,najafi2018sandpile}, self-avoiding walks~\cite{cheraghalizadeh2018self}, loop-erased random walks~\cite{daryaei2014loop}, the Ising model~\cite{cheraghalizadeh2018ising}. The case $p\ge p_c$ is considered in this paper, where the fluid moves through the largest percolating cluster.\\

\textbf{\textit{2-Ising Model}}, which has originally been developed for magnetic systems, but we use it to model the impermeable sites configuration. It is a good choice for this study since: (1) it is binary variable, (2) minimally makes the lattice correlated, (3) has a tuning parameter (artificial temperature $T$) which controls the correlations. The variables in this model are $s=+1$ and $s=-1$. The interactions in this model are short range (first neighbor). This model, in the zero magnetic field limit, is described by the following 
\begin{equation}
H =  - J\sum\limits_{\left\langle {ij} \right\rangle } {{s_i}{s_j}},
\end{equation}
where ${\left\langle {ij} \right\rangle }$ shows that the summation is over the nearest neighbor sites. The coupling constant $J$ identifies the type of interactions, we consider the ferromagnetic interaction ${J_{ij}} > 0 $. The correlations in this model is tuned by the (here artificial) temperature $T$. In two dimensions this model undergoes the magnetic phase transition (from para to ferro magnetic phase, $T>T_c$ and $T<T_c$ respectively)~\cite{baierlein1999thermal,gallavotti1999coexistence,cheraghalizadeh2018ising}. This phase transition is along with a percolation phase transition in two dimension, i.e. for $T<T_c$ a percolation geometric cluster is formed which comprised of connected sites with the same spin. We identify such clusters using the HK method~\cite{hoshen1976percolation}, and define the IP model on top of this cluster. The dynamical aspects of many statistical models have been already studied on the Ising-correlated lattices, like the sandpiles~\cite{cheraghalizadeh2017mapping,najafi2020geometry}, self-avoiding walks~\cite{cheraghalizadeh2018self}, loop-erased random walks~\cite{Cheraghalizadeh2019}, the Ising model~\cite{cheraghalizadeh2018ising}. In all cases a power-law behavior is seen for the \textit{exponents} in the vicinity of the critical point, which is called the ``secondary power-law'' behavior. To generate the Ising samples we used the Swendsen-Wang algorithm~\cite{swendsen1987nonuniversal} to avoid the critical slowing down problem. In this algorithm, instead of single spin flip, which is done in the Metropolis method, one flips a connected cluster of Fortuin-Kasteleyn (FK) clusters, which is a geometric connected cluster with same oriented spins for which the spins are connected with the probability $P_{\text{link}}=1-e^{-2/T}$. This approach is proved to be more efficient in the vicinity of the critical points~\cite{swendsen1987nonuniversal}.\\

\textbf{\textit{3-random coulomb potential (RCP)}}, which is used to make the interactions of the correlated lattice long-range. In fact, we consider a quenched correlated system through which the IP dynamics occurs. Let us first give a brief description of RCP production. Here we present a brief explanation of RCP which is an important model in the theory of probability and statistical mechanics. To describe the model, let us consider a 2D contentious system which is described in terms of a random field $\tilde{h}(\vec{x})$, where $\vec{x}$ represents the points in the system. The Edwards Wilkinson model, which in the stationary regime is a described by (and is a representation of) RCP, is defined via the following dynamical equation\\
\begin{equation}
{\partial _t}\tilde{h}(\vec x,t) = {\nabla ^2}\tilde{h}(\vec x,t) + \eta (\vec x,t),
\end{equation}
where $\eta (\vec x,t) $ is a space-time white noise with properties
\begin{equation}
\left\langle {\eta (\vec x,t)} \right\rangle  = 0\ ,\ \left\langle {\eta (\vec x,t)\eta (\vec x',t)} \right\rangle  = \xi {\delta ^2}(\vec x - \vec x')\delta (t - t') 
\end{equation}
in which $\xi  $ is the strength of the noise. The stationary phase is defined as the regime where the statistical average and correlations of $\tilde{r}$ become time-independent, i.e. ${\partial _t}\left\langle \tilde{h} \right\rangle  = 0 $. In this regime, the system become equivalent to a system described by the following time-independent equation
\begin{equation}
{\nabla ^2}\tilde{h}(\vec{x}) =  - \rho (\vec{x})
\label{Eq:GFF}
\end{equation}
where $\rho$ is a new normal distribution time-independent noise with the properties
\begin{equation}
\left\langle {\rho (\vec x')} \right\rangle  = 0,\ \left\langle {\rho (\vec x)\rho (\vec x')} \right\rangle  = {({n_i}a)^2}{\delta ^2}(\vec x - \vec x') 
\end{equation}
which represents the Poisson equation with the dielectric constant $\varepsilon  \equiv 1 $. In the above equation $a $ is the lattice constant and $ {n_i}$ is the density of Coulomb disorders. We use $r$ as the quenched correlated random number which realizes the pores quality for passing fluid, based on which the dynamics of the IP model is defined. To make the system similar to the two cases considered above, we normalize $\tilde{r}$ as 
\begin{equation}
h(\vec{x}) = \frac{1}{2}\left(\frac{\tilde{h}(\vec{x})}{\tilde{h}_{\max }} + 1\right)
\label{Eq:GFF2}
\end{equation}
where $\tilde{h}_{\max}\equiv\max\left\{\tilde{h}(\vec{x})\right\}_{\vec{x}\in \text{lattice}}$, so that $h(\vec{x})$ becomes correlated normal distribution variable in the range $[0,1]$. For generating samples we first distribute random charges ($\rho$ variable) throughout the square lattice and solve the Eq.~\ref{Eq:GFF} using the self-consistent iteration method. The RCP has often been employed as a model to be combined with many other dynamical models, like the percolation~\cite{cheraghalizadeh2018gaussian2} and Ising~\cite{cheraghalizadeh2018gaussian} models. For a good review of RCP model see~\cite{sheffield2007gaussian}.

\section{the numerical details and results}
In this section we present the results of IP dynamics on the short-range (percolation and Ising) as well as the long-range (RCP) correlated supports. The lattice sizes considered in this work are $L=32,64,128,256,512$, and we produced over $10^4$ samples for each system size and any control parameter ($T$ for the Ising and $p$ for the percolation). For the Ising model we considered the temperatures $T=1.8,2.0,2.1,2.2,2.26918$, and for the percolation case we have taken into account the occupation probabilities $p=0.59275,0.6,0.65,0.75,1.0$. We implemented the model in two separate geometries: $L\times L$ square lattice with free boundary conditions, and the cylinder geometry which is used to extract the fractal dimension of the largest hole.\\
For generating the RCP samples, we used self-consistent iteration method. In the square lattice we consider open boundary conditions in both directions, whereas in the cylinder geometry we consider the open boundary conditions in one free direction.

\subsubsection{IP in percolation- and Ising-correlated support}
\begin{figure*}
	\begin{subfigure}{0.47\textwidth}\includegraphics[width=\textwidth]{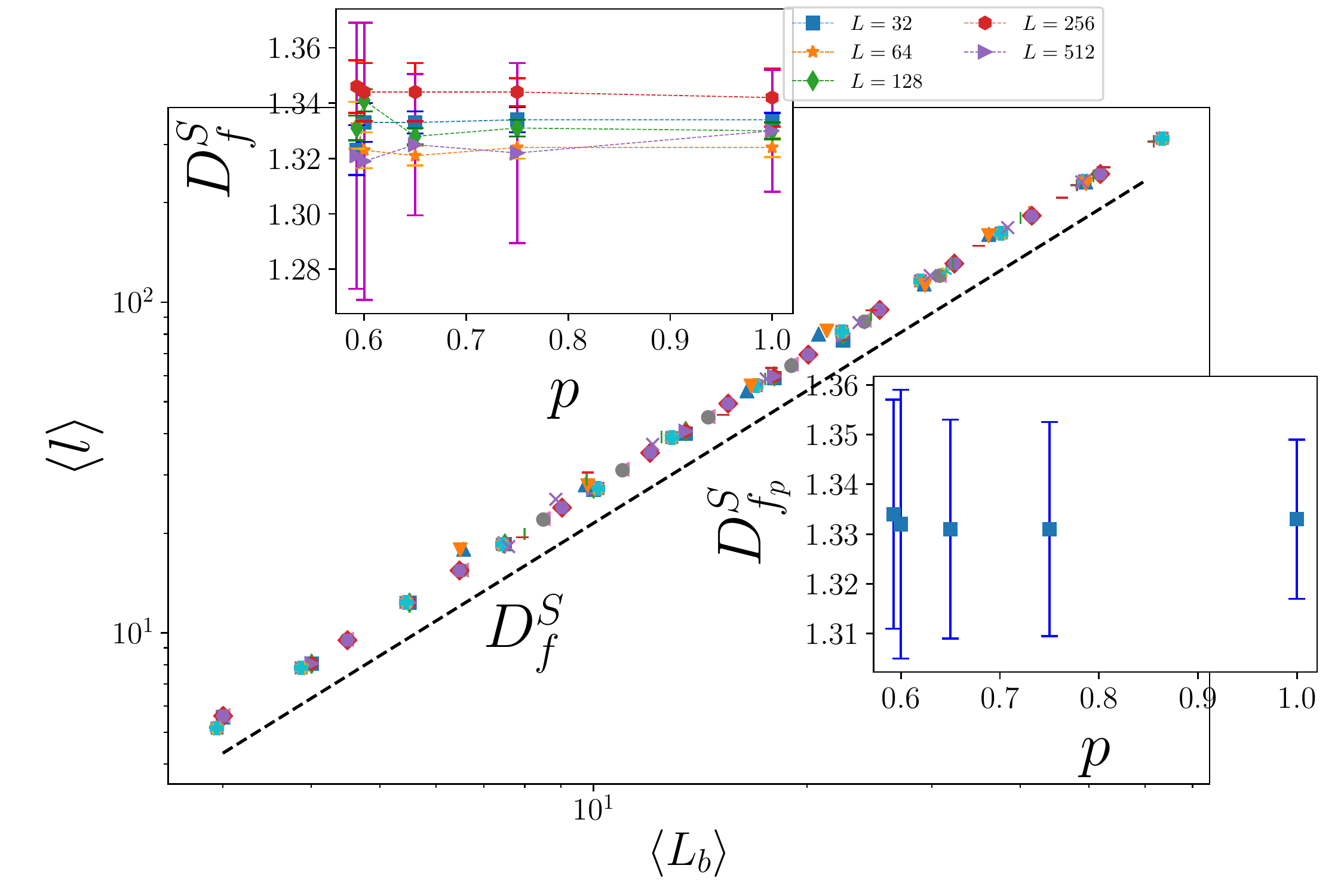}
		\caption{}
		\label{perbox}
	\end{subfigure}
	\begin{subfigure}{0.47\textwidth}\includegraphics[width=\textwidth]{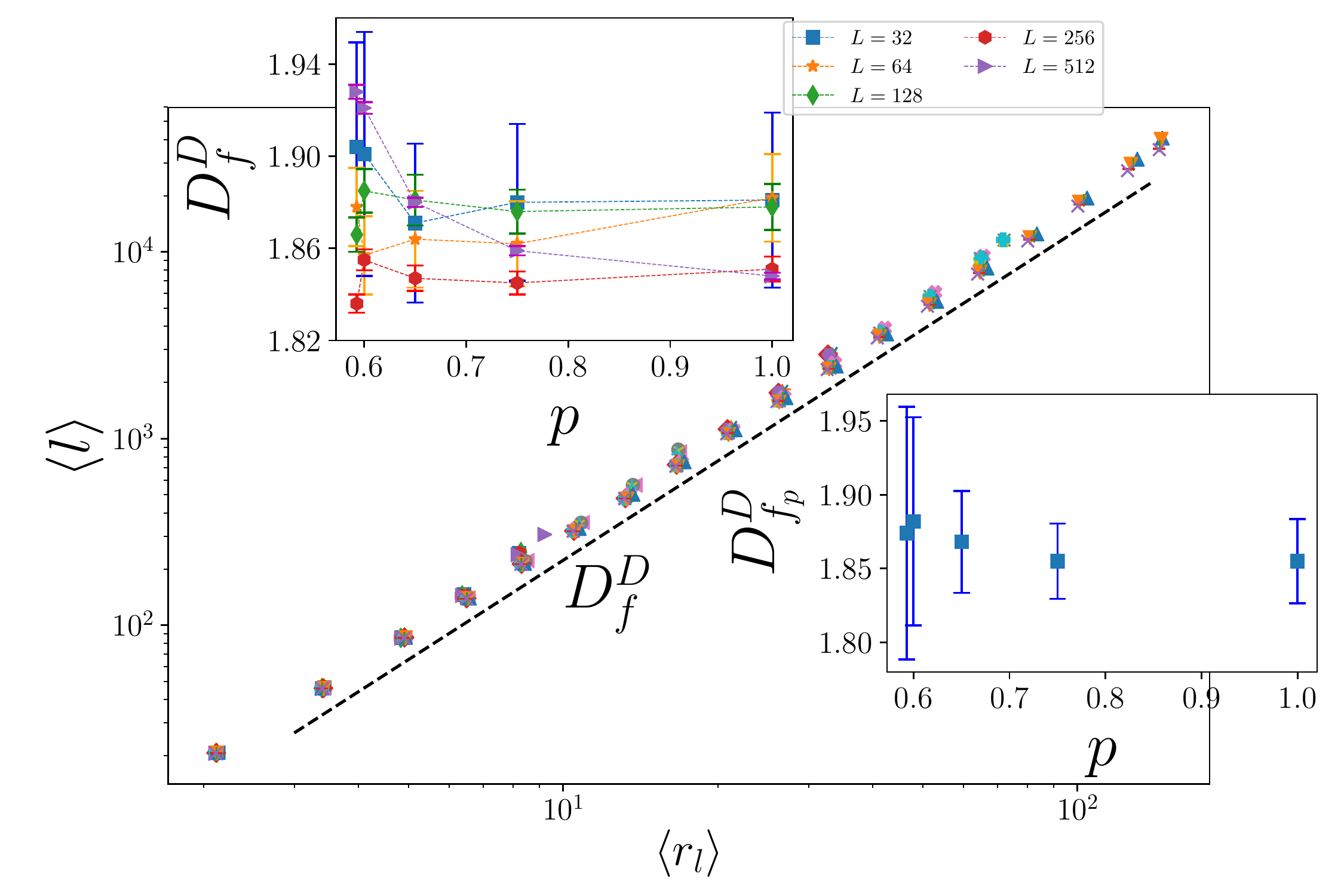}
		\caption{}
		\label{rl-lper}
	\end{subfigure}
    	\begin{subfigure}{0.47\textwidth}\includegraphics[width=\textwidth]{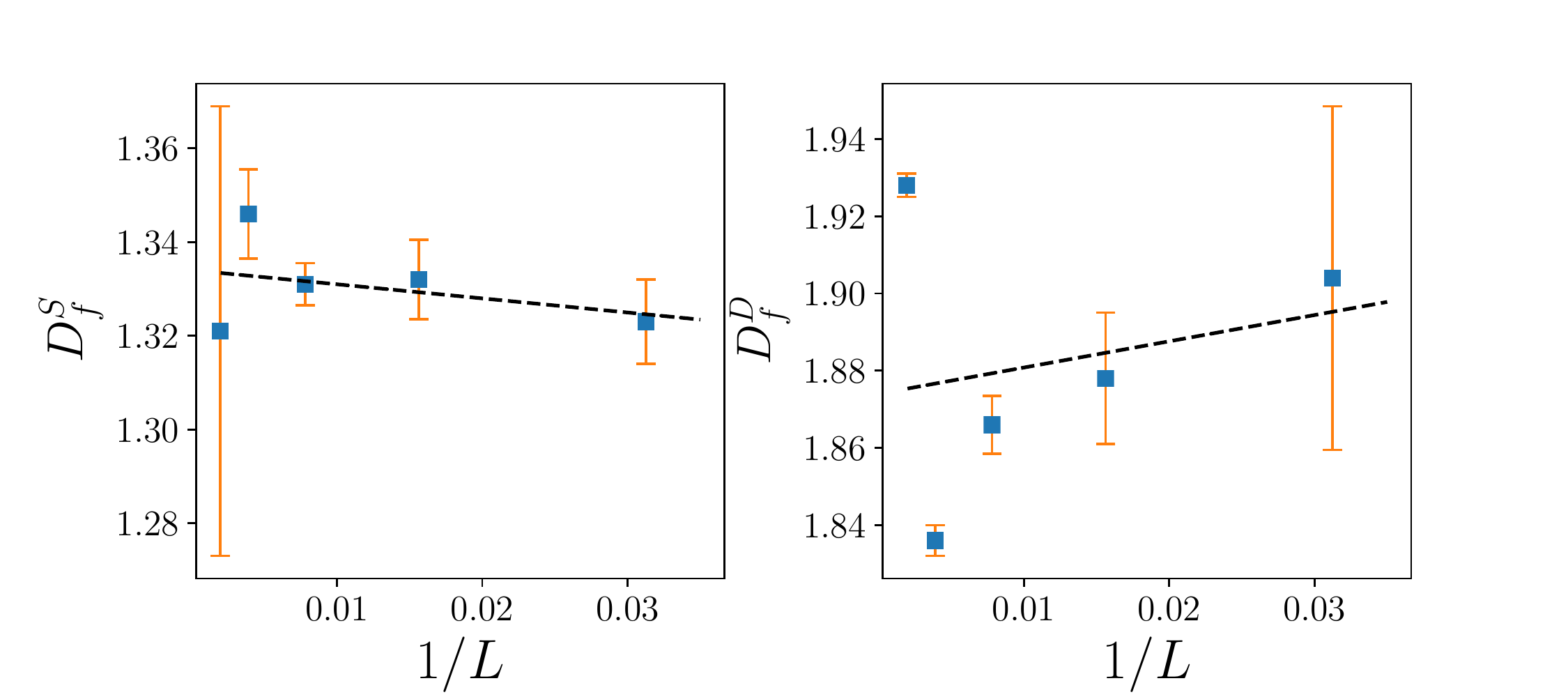}
    	\caption{}
    	\label{slope}
    \end{subfigure}
	\caption{(Color Online) The numerical results for percolation background (a) The numerical results of the fractal dimension, which is the slope of the $l-L_{b}$ graph in the log-log plot. (b) log-log plot of $l-r_{l}$ graph, that slope is the fractal dimension. (c) $D_{f}^{S} $ and $D_{f}^{D} $ in term of $1/L$ for $p=p_{c}$($p_{c}=0.59275$).}
	\label{percolation2}
\end{figure*}

presumably the most important quantity in IP is the fractal dimension of the boundaries of the largest hole, which is $\frac{4}{3}$ for the regular IP. For the IP-percolation the results are shown in Fig.~\ref{percolation2} for $p>p_c$ and all of the $L$ that we use in simulation. The $L$-dependence of the exponent ($D_f^S(L)$) is shown in the upper inset, whereas the lower inset shows the exponent in the thermodynamic limit ($D_f^S(\infty)$), which is obtained using the extrapolation relation
\begin{equation}
D_f^S(L)=D_f^S(\infty)+\frac{A}{ L}
\label{Eq:FSS}
\end{equation}
where $A$ is a non-universal proportionality constant. We see that this exponent ($D_f^S$) does not significantly run with $p$ being fixed on the theoretical prediction $1.32\pm 0.03$, even in the vicinity of the critical point for which the error bars are higher. In Fig.~\ref{rl-lper} the DFD is reported, for which $D_{f}^D$ is robust against $p$, and is fixed to $1.9\pm 0.2$.\\

\begin{figure*}
	\begin{subfigure}{0.47\textwidth}\includegraphics[width=\textwidth]{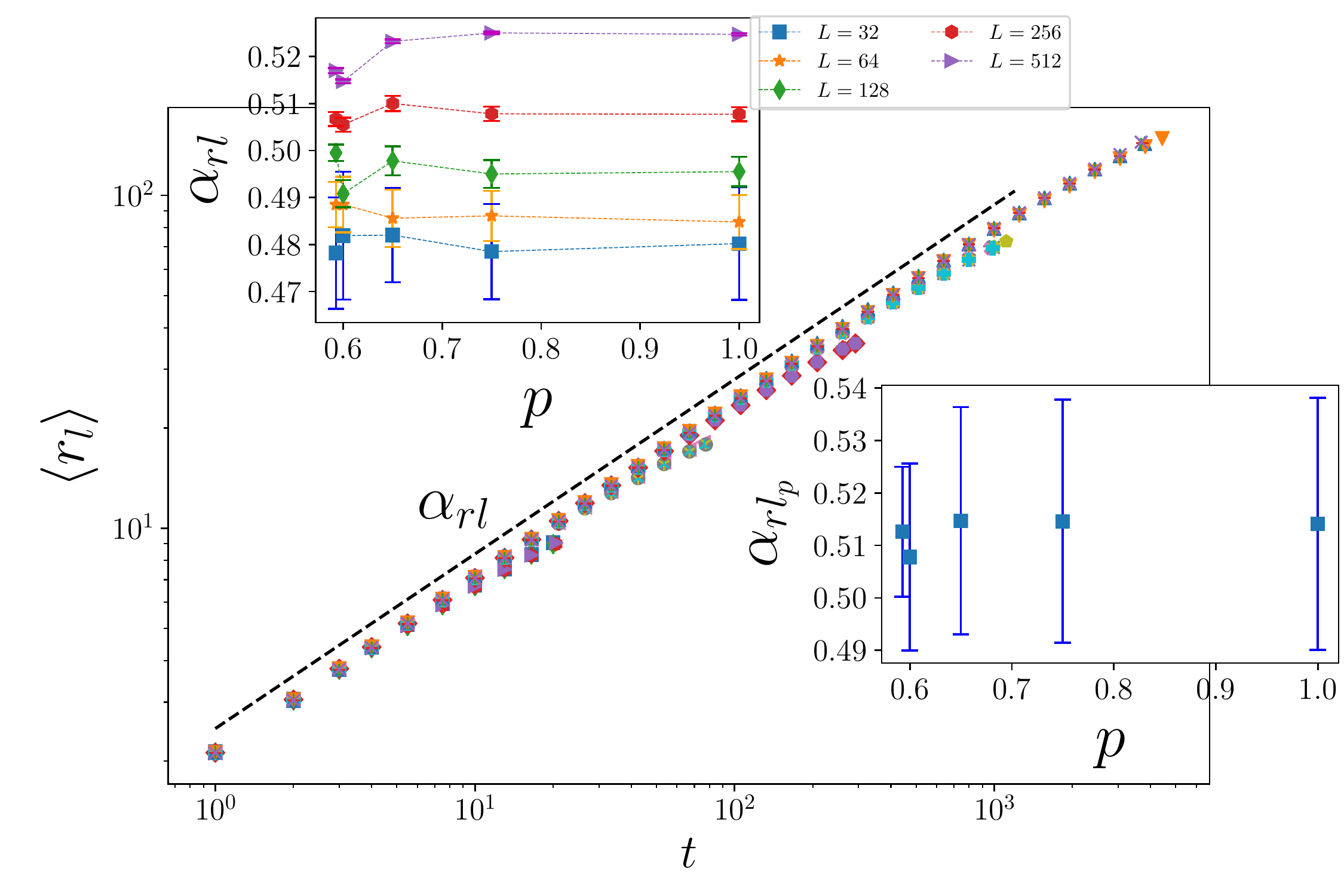}
		\caption{}
		\label{fig:trlper}
	\end{subfigure}
	\begin{subfigure}{0.47\textwidth}\includegraphics[width=\textwidth]{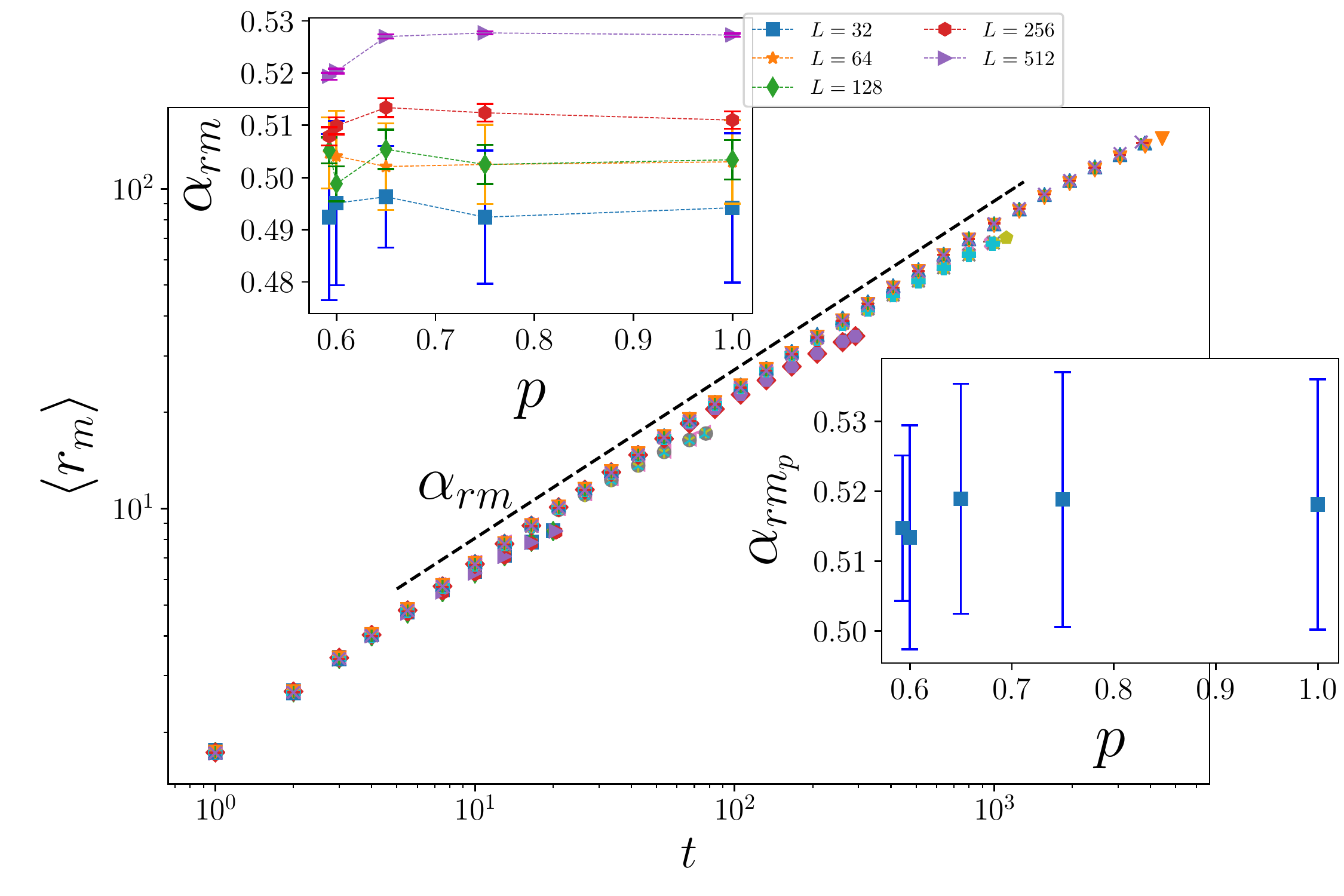}
		\caption{}
		\label{fig:trmper}
	\end{subfigure}
	\begin{subfigure}{0.47\textwidth}\includegraphics[width=\textwidth]{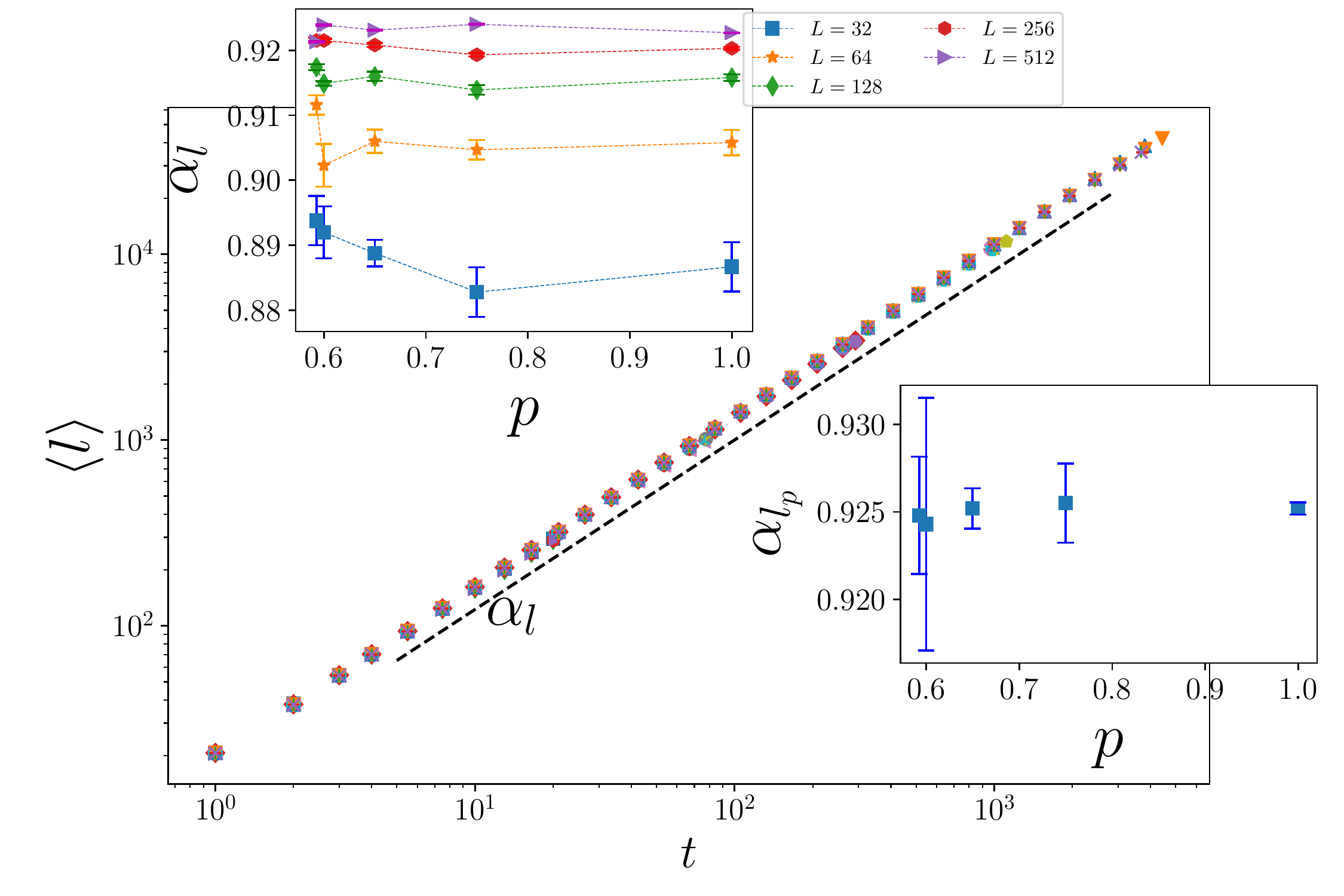}
		\caption{}
		\label{fig:tlper}
	\end{subfigure}
	\begin{subfigure}{0.47\textwidth}\includegraphics[width=\textwidth]{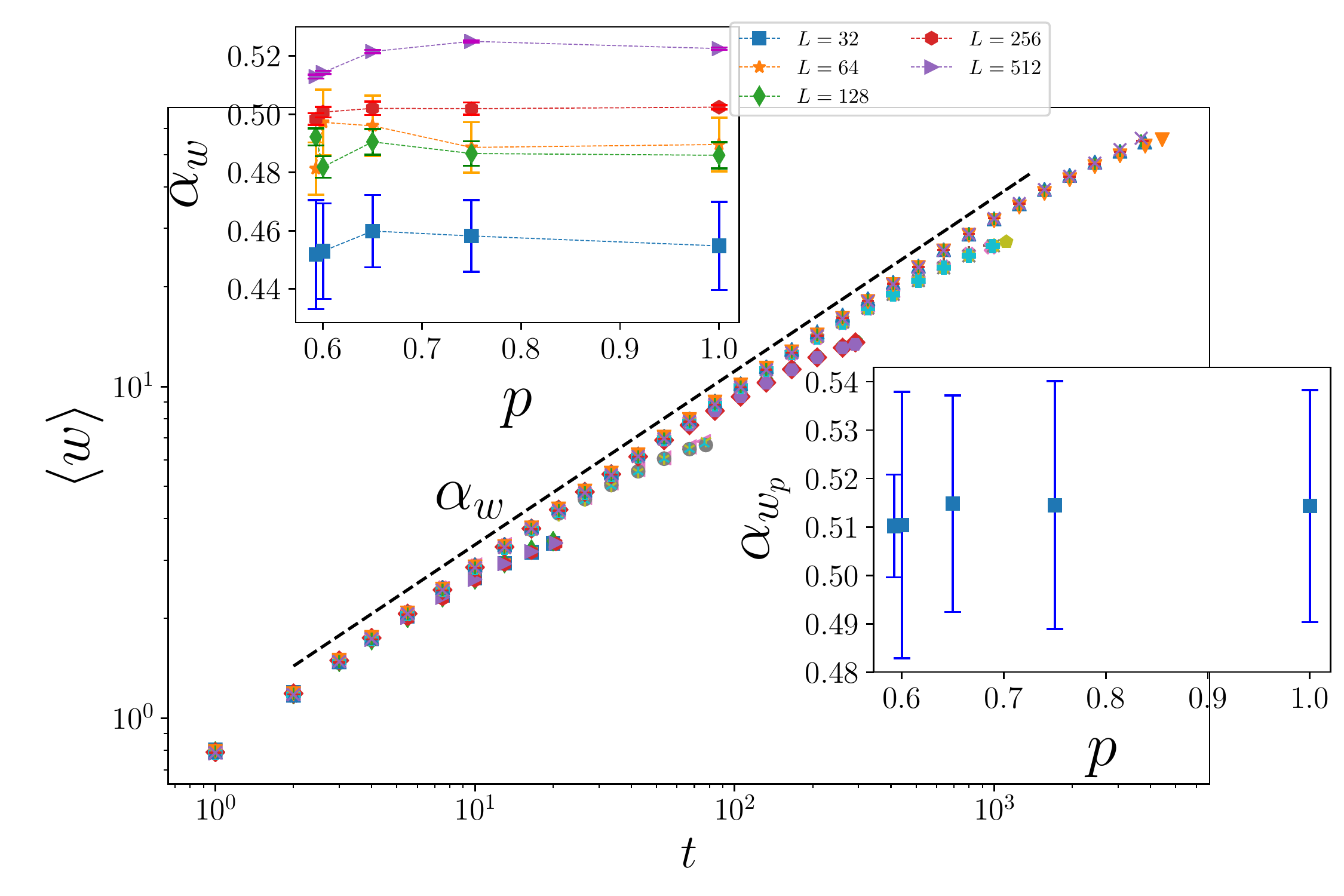}
		\caption{}
		\label{fig:twper}
	\end{subfigure}
	\caption{(Color Online) (a) The time dependence of the average of $l$. (b)The time dependence of the average of $r_{l}$. (c)The time dependence of the average of $r_{m}$. (d)The time dependence of the average of $w$.}
	\label{Fig:percolation}
\end{figure*}

To investigate this more deeply and observe if this robustness is general, we have considered the time dependence of $l$, $r_l$, $r_m$ and $w$ which are represented in Fig.~\ref{Fig:percolation}, showing that they behave in a power-law fashion. From Figs.~\ref{fig:trlper} and~\ref{fig:trmper} we see that the system is in the normal diffusion regime, which is detected by the $r-t$ exponent $\alpha_r=0.5\pm 0.05$ for all $p$ values. The exponent $\alpha_l$ should be compatible with the scaling argument 
\begin{equation}
l\propto t^{\alpha_l}\propto \left[t^{\alpha_r}\right]^{\frac{\alpha_l}{\alpha_r}}\propto r_l^{\frac{\alpha_l}{\alpha_r}},
\end{equation}
so that $\alpha_l=D_f^D\alpha_r=0.9\pm 0.05$, which is compatible with the Fig.~\ref{fig:tlper}.  \\
In the conventional growth models, the roughness shows power-law behavior in early time stages, and enters a stationary regime after a crossover time. In the latter regime the absolute value of roughness varies with the system size in a power-law form~\cite{kondev2000nonlinear,barabasi1995fractal}. In our model, shown in Fig.~\ref{fig:twper}, the roughness exhibits a power-law behavior, and before entering the stationary phase, the process finishes, no matter what the system size is. The exponent of this power-law behavior is $\alpha_w=0.45\pm 0.05$, which is almost constant for all $p$ values.

\begin{figure*}
	\begin{subfigure}{0.47\textwidth}\includegraphics[width=\textwidth]{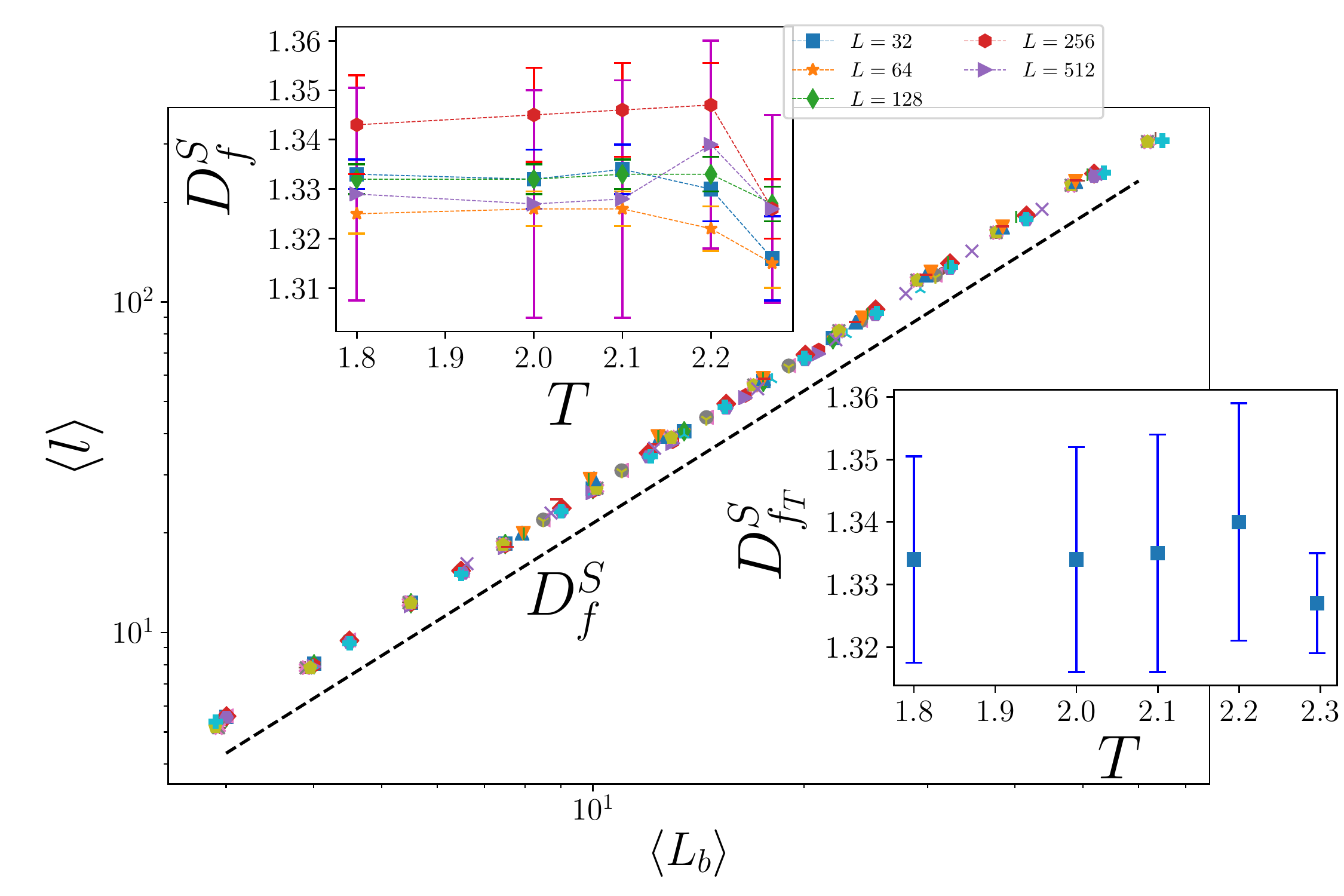}
		\caption{}
		\label{isingbox}
	\end{subfigure}
	\begin{subfigure}{0.47\textwidth}\includegraphics[width=\textwidth]{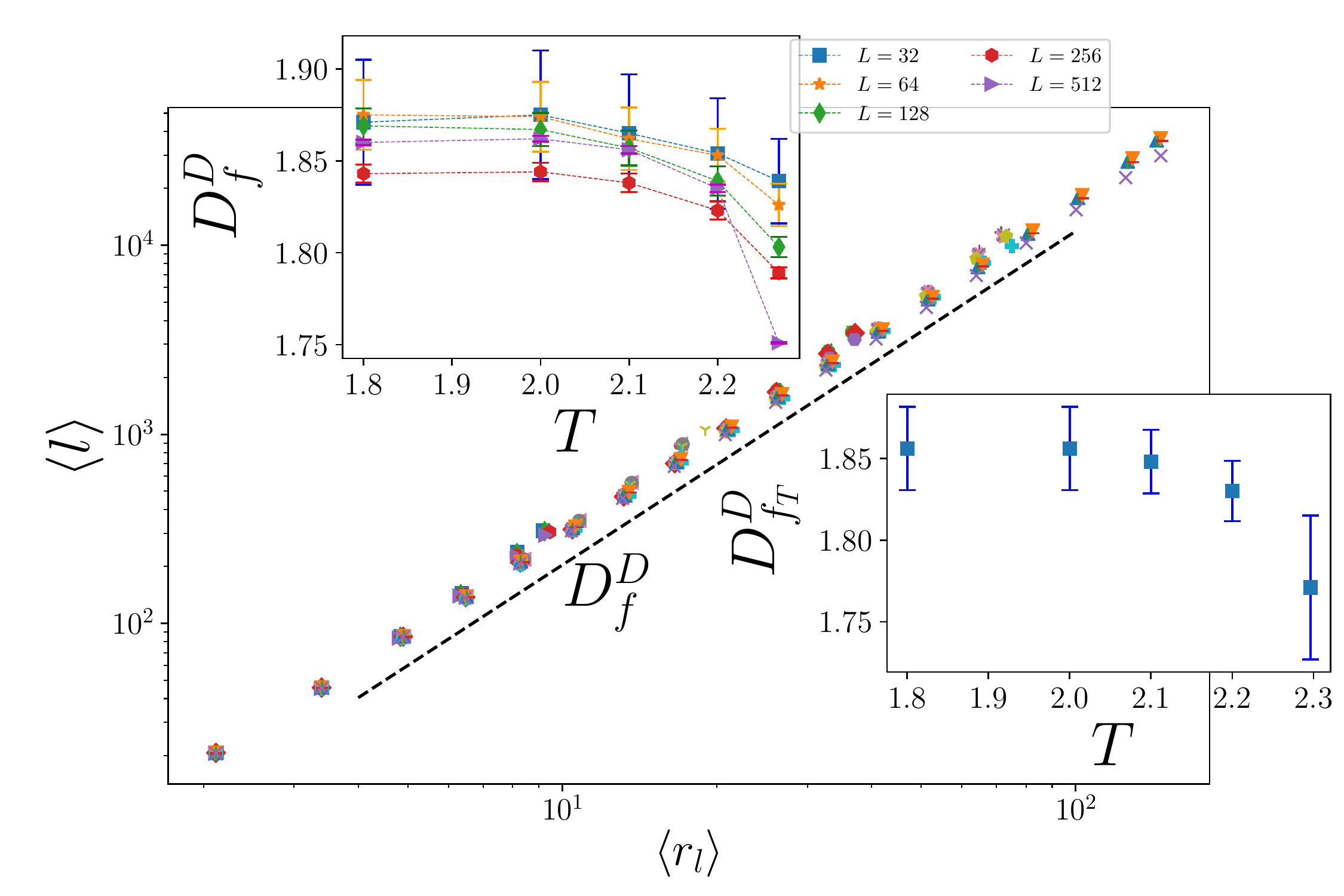}
		\caption{}
		\label{rl-lising}
	\end{subfigure}
		\begin{subfigure}{0.47\textwidth}\includegraphics[width=\textwidth]{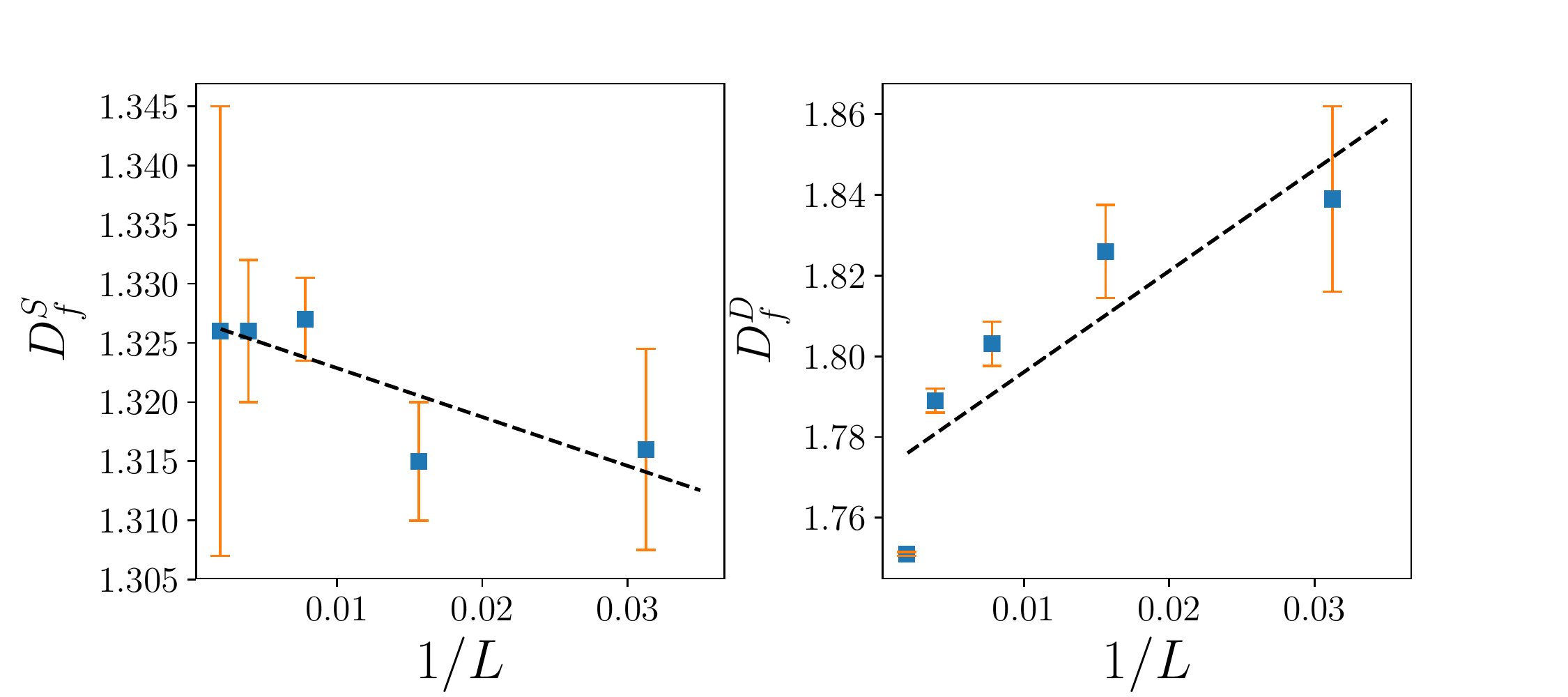}
		\caption{}
		\label{slope}
	\end{subfigure}
	\caption{(Color Online) The numerical results for ising background (a) The numerical results of the fractal dimension, which is the slope of the $l-L_{b}$ graph in the log-log plot. (b) log-log plot of $l-r_{l}$ graph, that slope is the fractal dimension. (c) $D_{f}^{S} $ and $D_{f}^{D} $ in term of $1/L$ for $T=T_{c}$($T_{c}=2.26918$).}
	\label{fig:ising}
\end{figure*}

\begin{figure*}
	\begin{subfigure}{0.47\textwidth}\includegraphics[width=\textwidth]{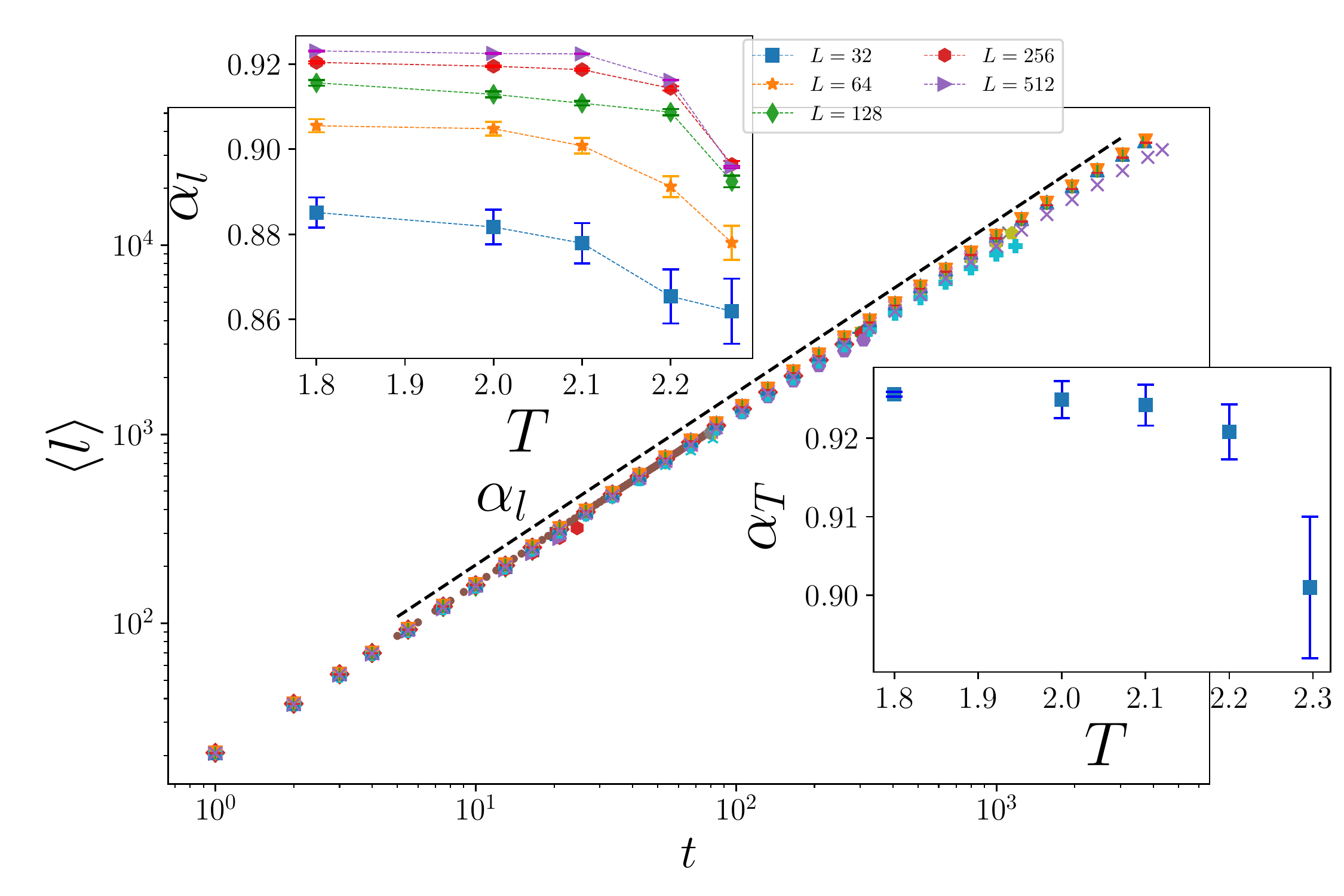}
		\caption{}
		\label{fig:tlising}
	\end{subfigure}
	\begin{subfigure}{0.47\textwidth}\includegraphics[width=\textwidth]{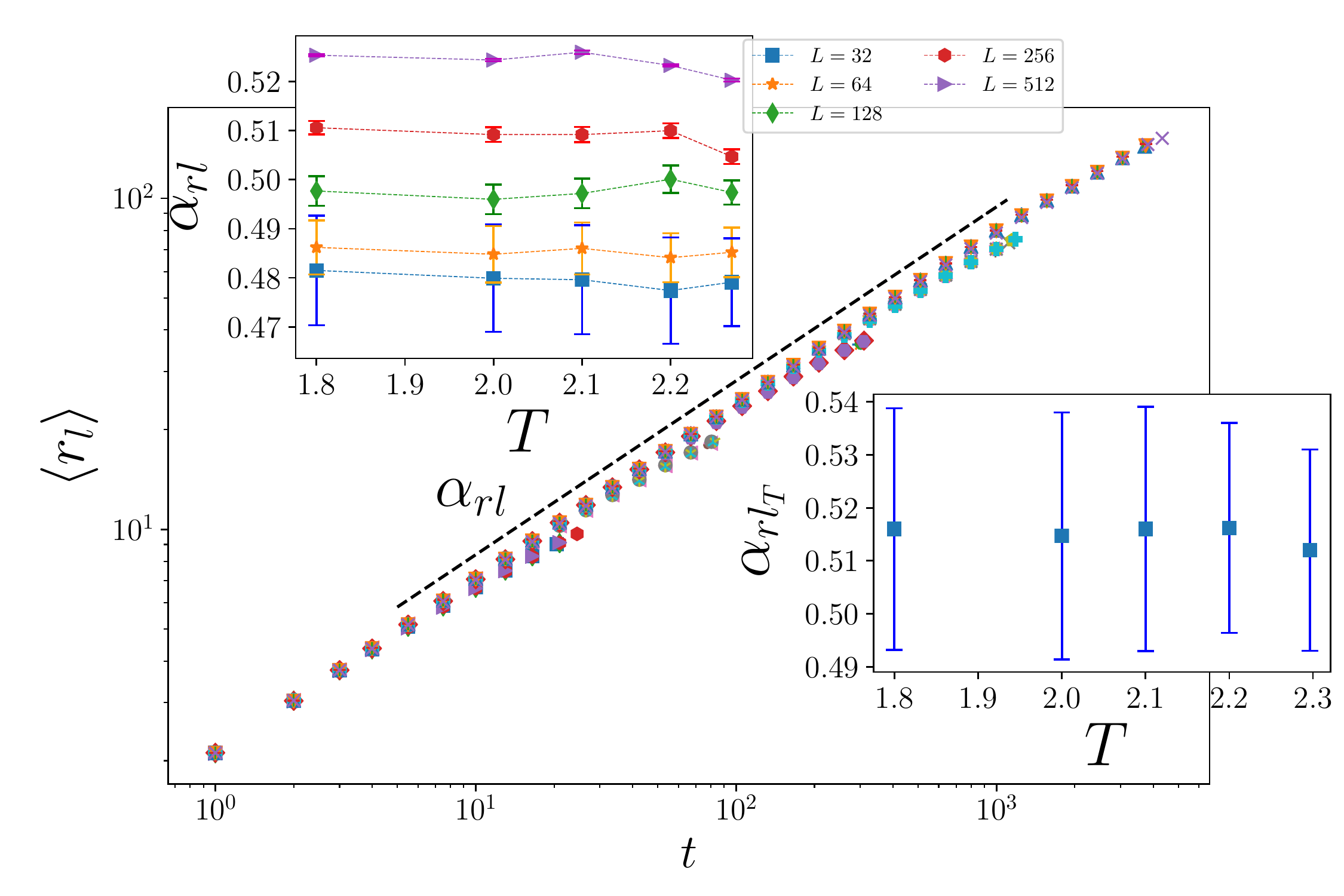}
		\caption{}
		\label{fig:trlising}
	\end{subfigure}
	\begin{subfigure}{0.47\textwidth}\includegraphics[width=\textwidth]{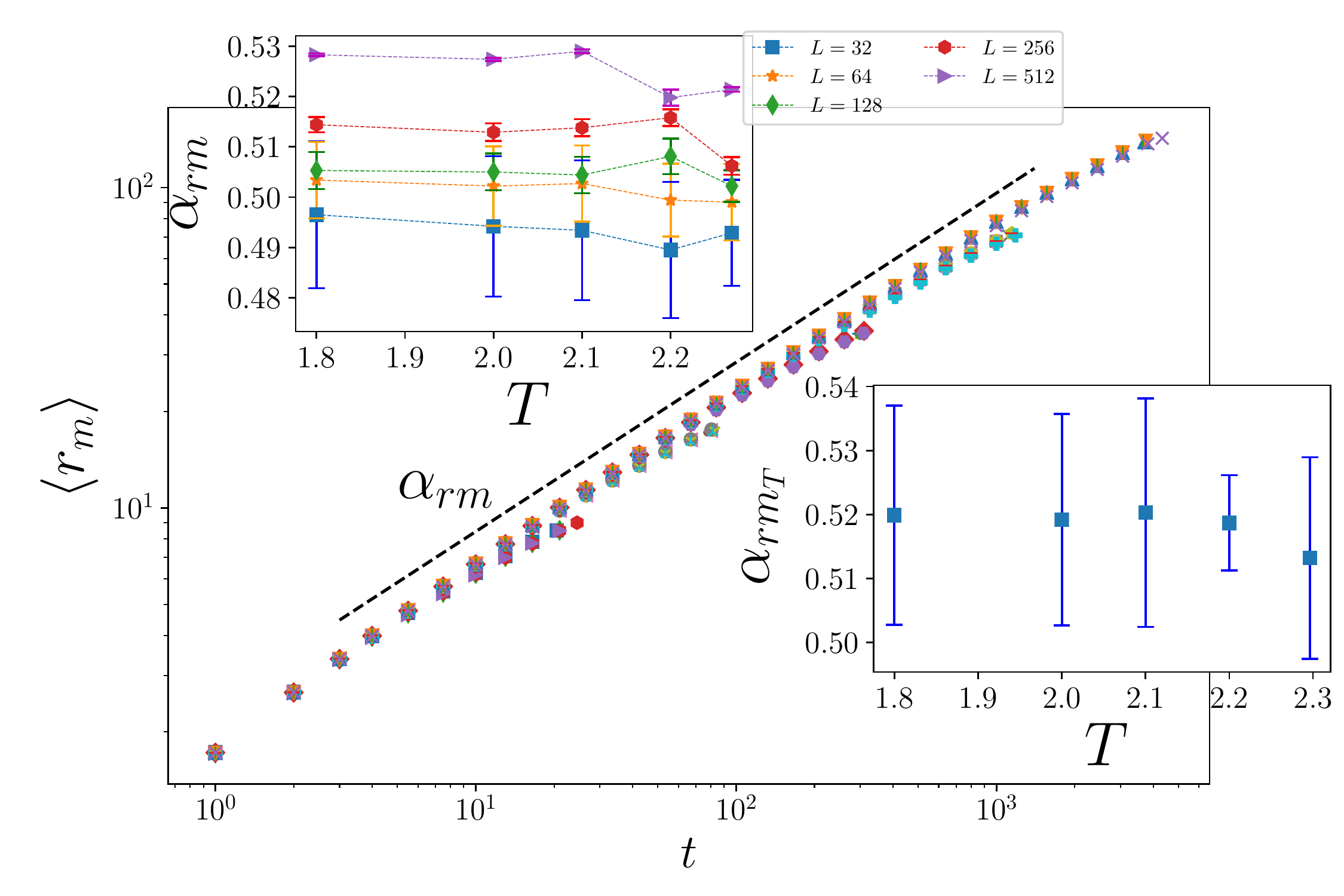}
		\caption{}
		\label{fig:trmising}
	\end{subfigure}
	\begin{subfigure}{0.47\textwidth}\includegraphics[width=\textwidth]{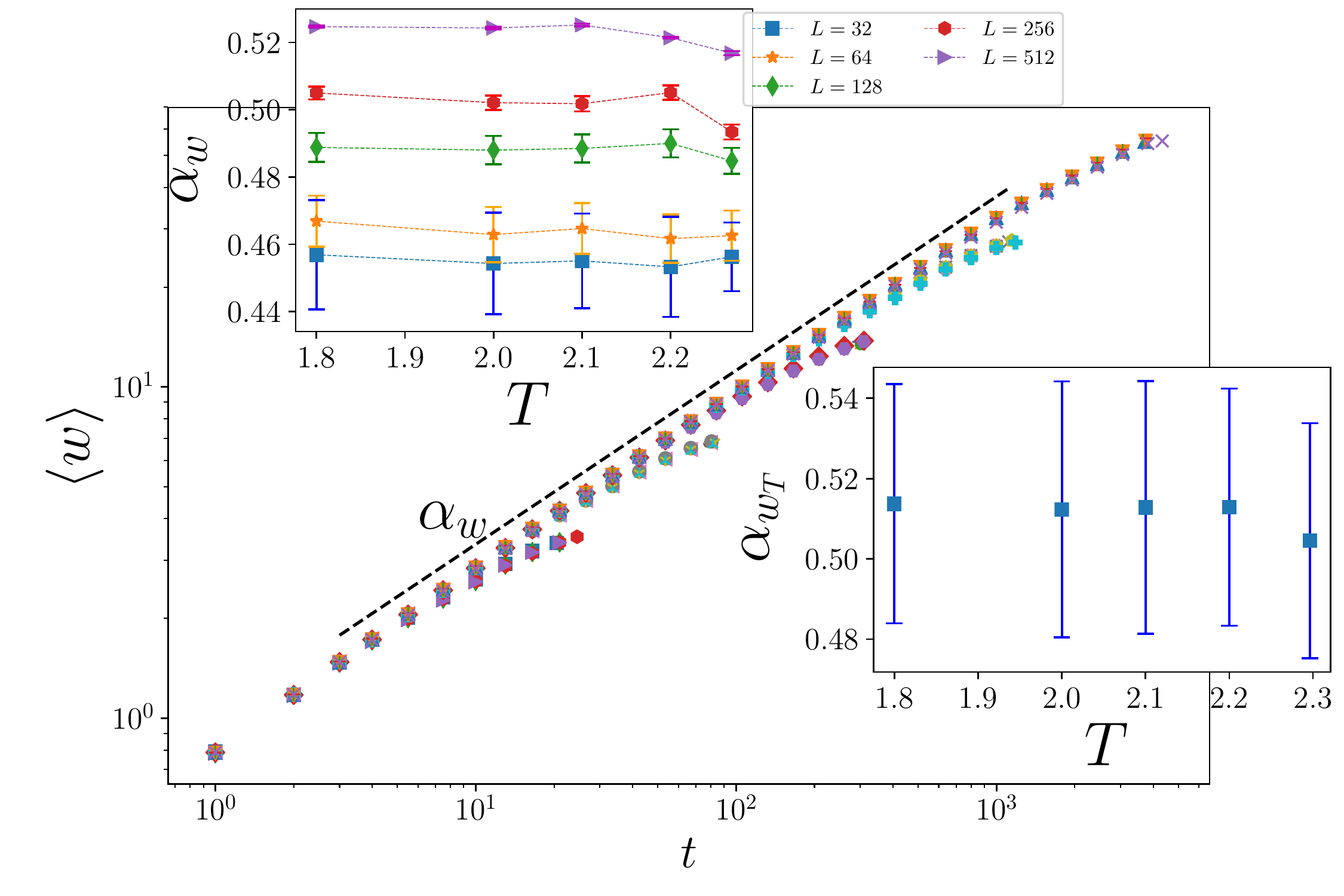}
		\caption{}
		\label{fig:twising}
	\end{subfigure}
	\caption{(Color Online) (a) The time dependence of the average of $l$. (b)The time dependence of the average of $r_{l}$. (c)The time dependence of the average of $r_{m}$. (d)The time dependence of the average of $w$.}
	\label{Eq:isingDynamic}
\end{figure*}

Interestingly we have observed that the correlations due to the short interactions in the Ising model does not change this behavior. The same analysis has been carried out in Figs.~\ref{fig:ising} and~\ref{Eq:isingDynamic}. This time, the exponents show robust behavior against $T<T_c$, which controls the range of correlations. The exponents do not significantly change even in the vicinity of the critical temperature. In Figs.~\ref{isingbox} and~\ref{rl-lising}, although the estimated fractal dimensions bind downwards in the vicinity $T_c$, they lie within the error bars of the estimated values at the lower temperatures in the thermodynamic limit ($L\rightarrow\infty$ shown in the lower insets, which has been obtained using the extrapolation relation Eq.~\ref{Eq:FSS}). The extrapolated exponents (thermodynamic limit, shown in the lower insets) reported in Figs.~\ref{fig:tlising}, \ref{fig:trlising}, \ref{fig:trmising}, and \ref{fig:twising} show more robust behavior.\\

Our calculations show that the introduction of uncorrelated (percolation) and correlated (Ising) lattice imperfections with short range interactions does not change the universality class of the IP model, i.e. it is an irrelevant perturbation for IP class. 

\subsubsection{IP on random coulomb potential (RCP)}-correlated support
The fact that the critical behaviors of IP for $p>p_c$ and $T<T_c$ are similar to the IP model in a perfect support is expected since in these intervals the properties of the host media is identical to the perfect (regular) support in large scales. The only case where the model has a chance to behave in a different way is right at the critical points, since in this case the host is self-similar and the features of the it repeats by re-scaling the space. The analysis in the previous section showed however that the properties of the model does not change even in the critical points. In RCP (and generally scale-invariant rough surfaces with scale-invariant Lagrangian) the system is always self-similar, and consequently the correlations are power-law having no referred scale. The difference with the two studied cases (percolation- and Ising-correlated supports) at the critical point is that, here in addition to the long-ranged correlations, the \textit{interactions} are also long range (Coulomb interactions). \\

In this section the \textit{pores quality} are supposed to be described by the relation~\ref{Eq:GFF2} whose correlations are proved to be~\cite{cheraghalizadeh2018gaussian,cheraghalizadeh2018gaussian2}
\begin{equation}
\left\langle h(\vec{x}+\vec{x}_0)h(\vec{x})\right\rangle\propto\log|x_0| 
\end{equation}
which is long range. This model is equivalent to scale invariant rough surface with zero roughness exponent. The properties of some dynamical models on (RCP)-correlated supports are already done~\cite{cheraghalizadeh2018gaussian,cheraghalizadeh2018gaussian2}. \\
\begin{figure*}
	\begin{subfigure}{0.47\textwidth}\includegraphics[width=\textwidth]{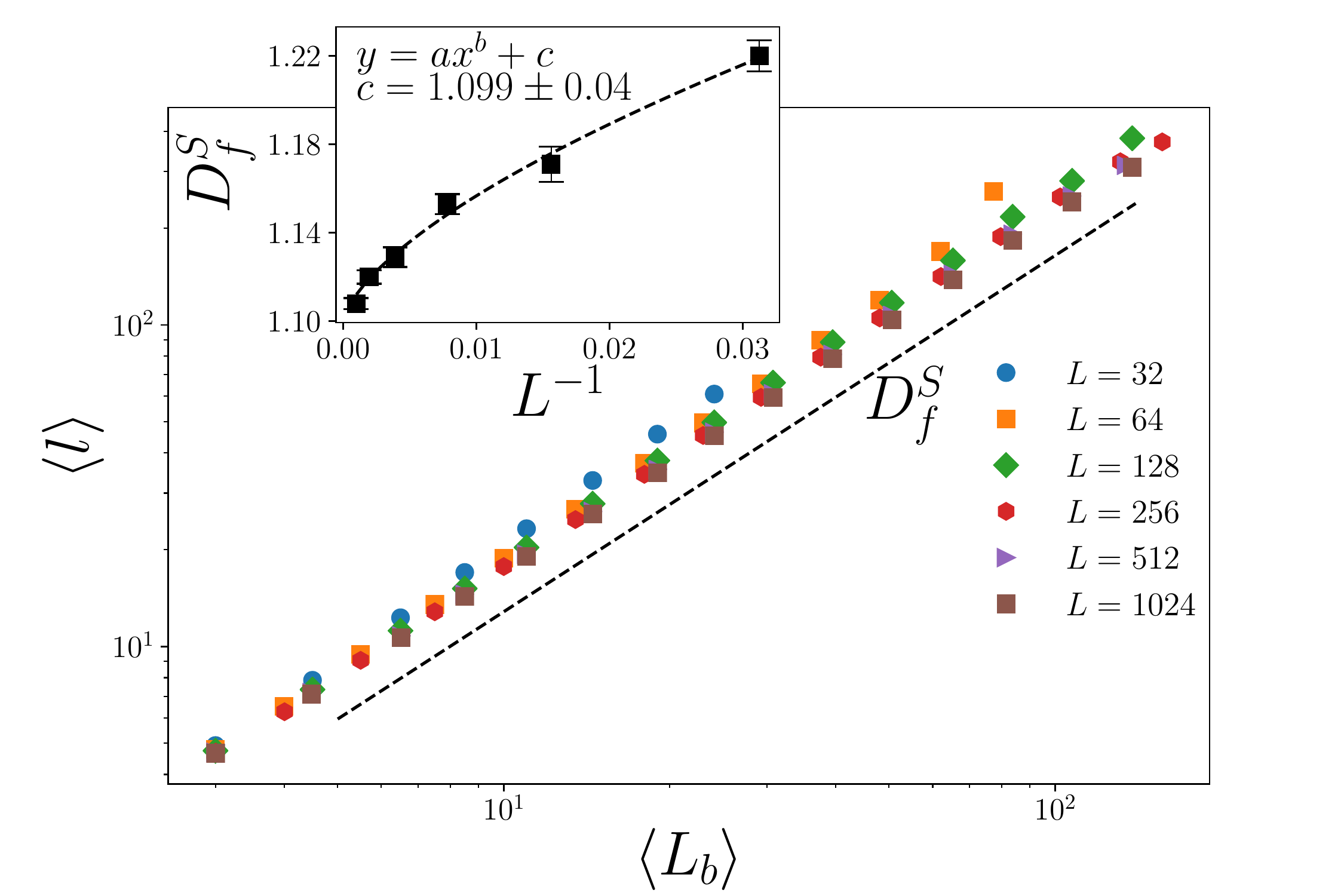}
		\caption{}
		\label{Eq:boxgffn}
	\end{subfigure}
	\begin{subfigure}{0.47\textwidth}\includegraphics[width=\textwidth]{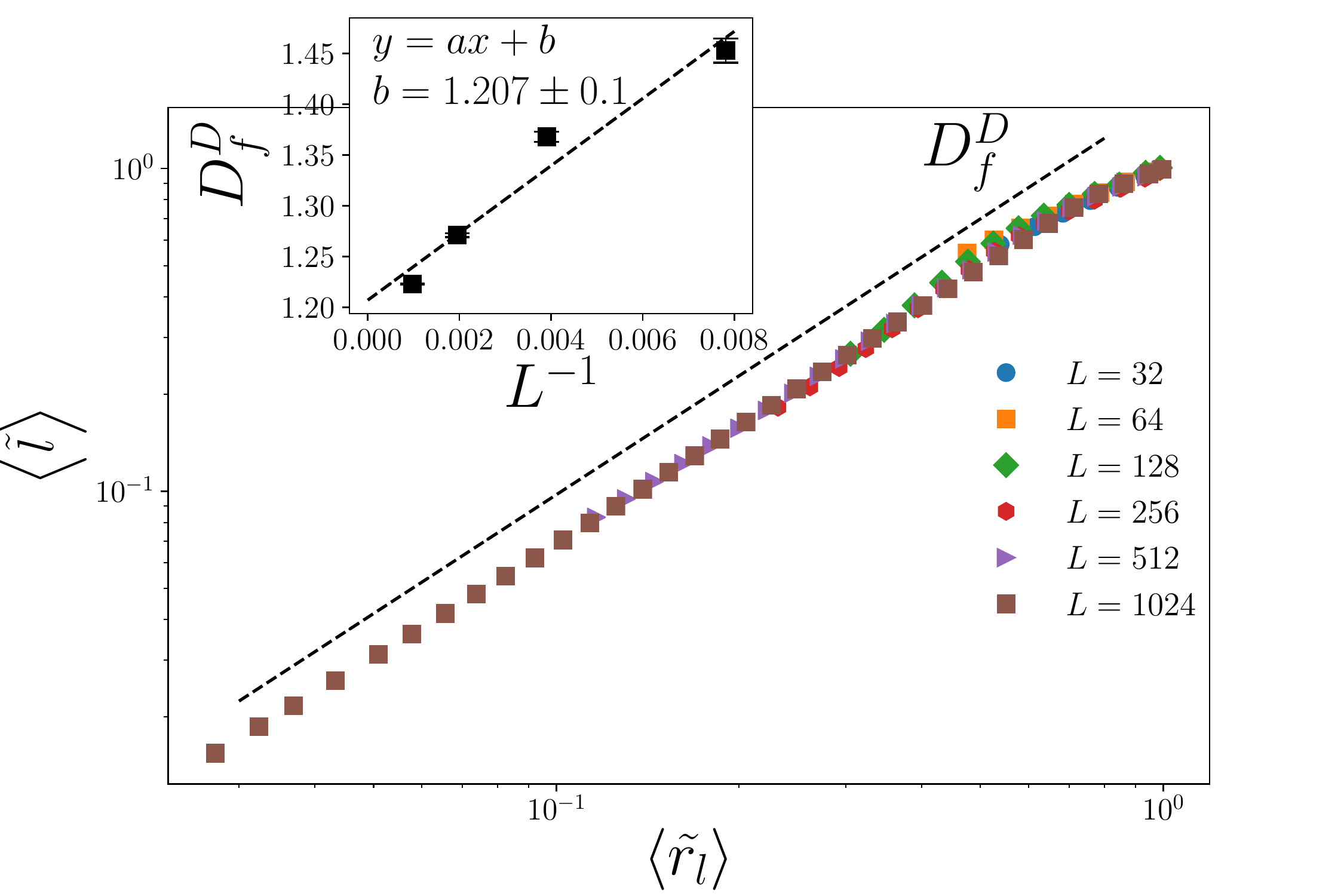}
		\caption{}
		\label{Eq:lrlgffn}
	\end{subfigure}
	\caption{(Color Online) The numerical results for random coulomb potential (RCP) background (a) The numerical results of the fractal dimension, which is the slope of the $l-L_{b}$ graph in the log-log plot. (b) log-log plot of $l-r_{l}$ graph, that slope is the fractal dimension.(Note that in all these insets, $x$ represents the horizontal axis and $y$ represents the vertical axis.)}
	\label{bdgff}
\end{figure*}
Here consider IP in a support where $r$ quantity is selected from the solution of Eq.~\ref{Eq:GFF2}. In this case the critical properties of IP changes significantly. Figure~\ref{Eq:boxgffn} shows that the SFD in the thermodynamic limit is $D_f^S(L\rightarrow\infty)=1.099\pm 0.040$ (see the inset). The dynamic fractal dimension is $D_f^D(L\rightarrow\infty)=1.207\pm0.1$. In this figure, we have used normalization to one, where the numbers on each axis are divided by their maximum (end point). Note that each graph including $\tilde{x} $ or $\tilde{y} $ symbol is normalized to one. Both of these exponents are significantly different from the ordinary IP model. This leads us to the conclusion the not only the long-range correlations are necessary to bring the system out of the IP universality class, but also the long-range interactions are necessary.\\

To be more precise, let us consider the dynamical aspects of the model. We have observed three distinct temporal regimes in the system. For small enough times (power-law regime where $t<t_1^*$) the observables show power-law behavior with time, whereas for intermediate times (logarithmic regime where $t_1^*<t<t_2^*$) it varies with the logarithm of time, and for long times (linear regime where $t>t_2^*$) they change linearly with time with a slope tending to zero in the thermodynamic limit. 
 \begin{figure*}
	\begin{subfigure}{0.45\textwidth}\includegraphics[width=\textwidth]{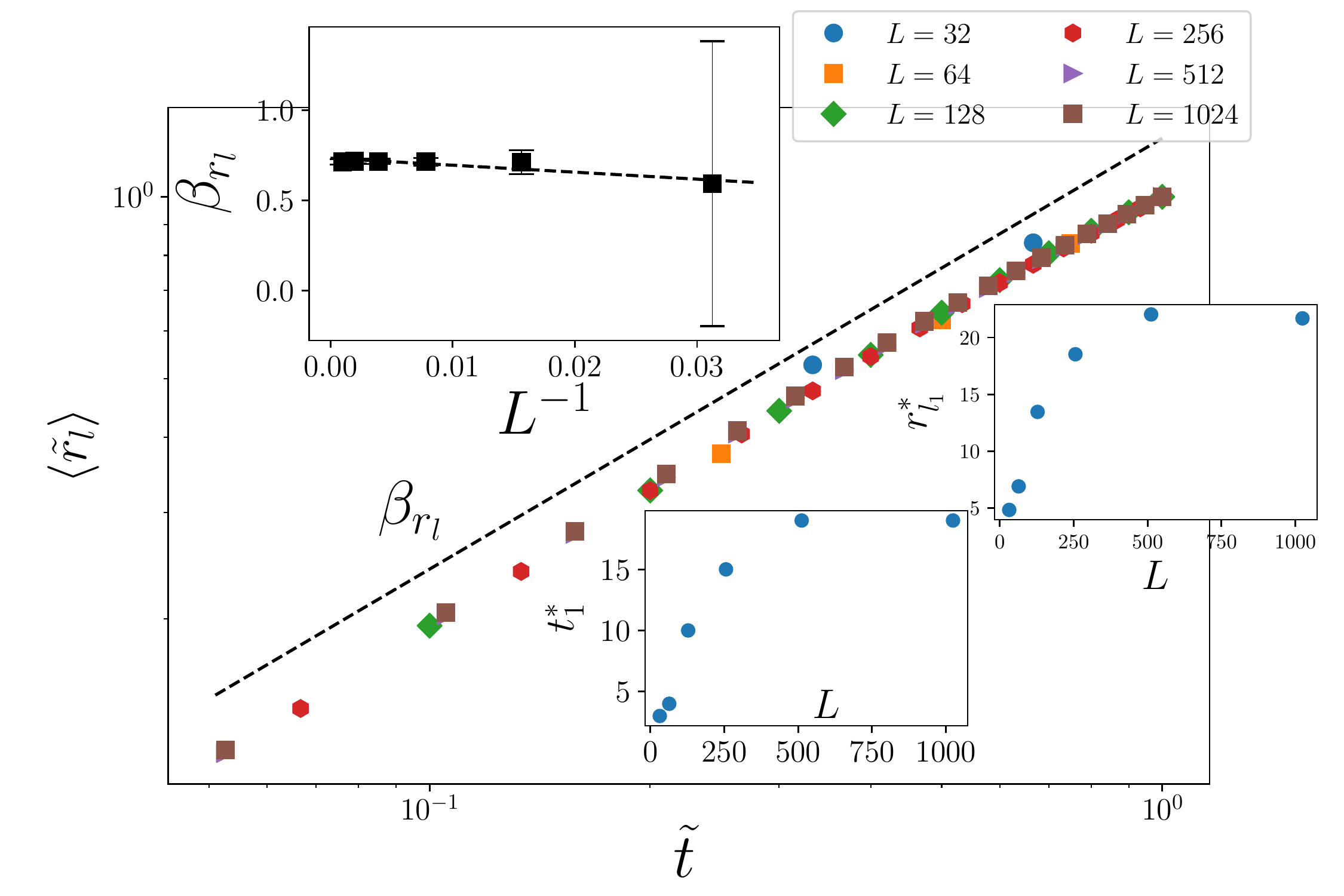}
		\caption{}
		\label{Fig:t_rlsmallnormalize}
	\end{subfigure}
	\begin{subfigure}{0.45\textwidth}\includegraphics[width=\textwidth]{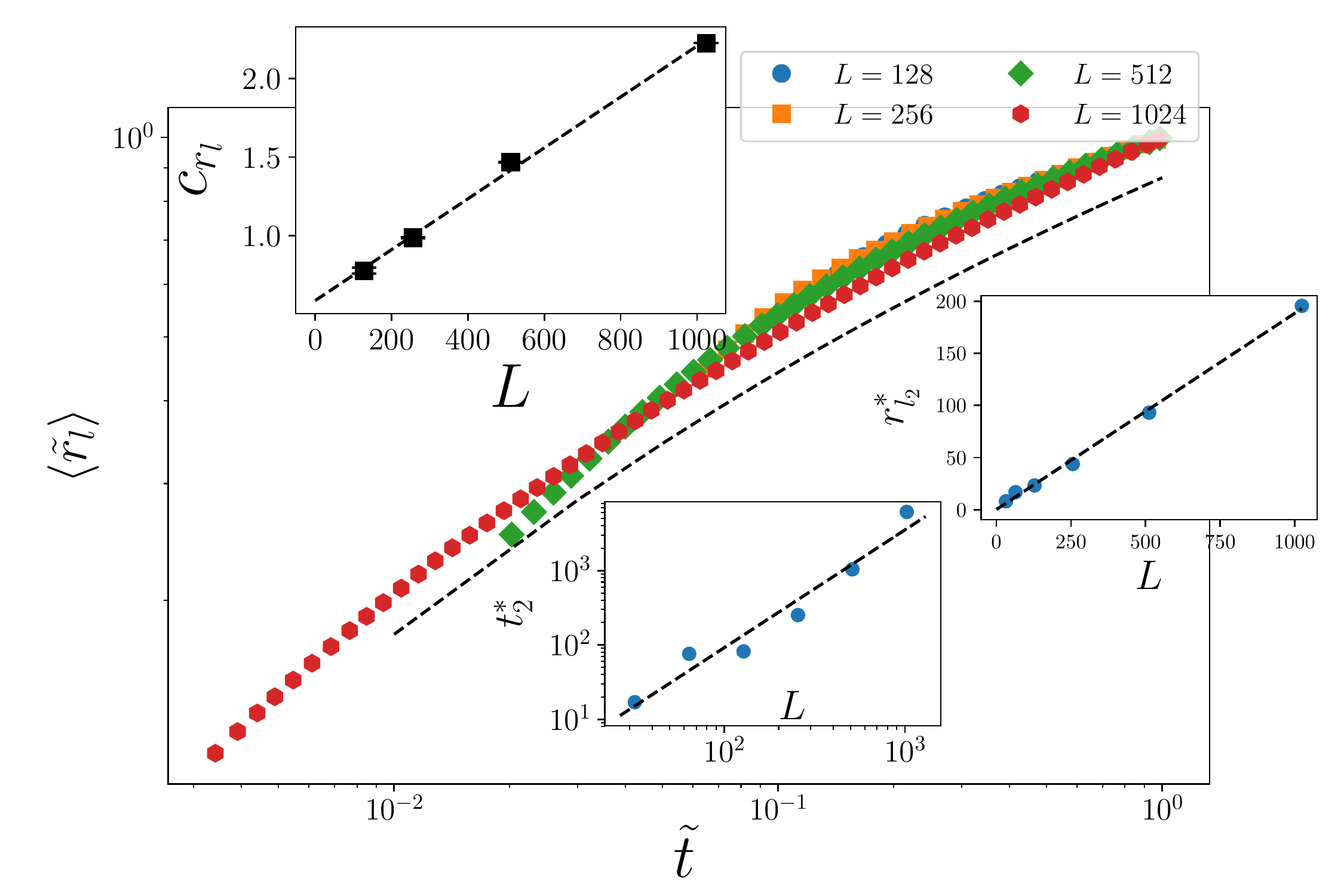}
		\caption{}
		\label{Fig:t_rlmiddle}
	\end{subfigure}
	\begin{subfigure}{0.45\textwidth}\includegraphics[width=\textwidth]{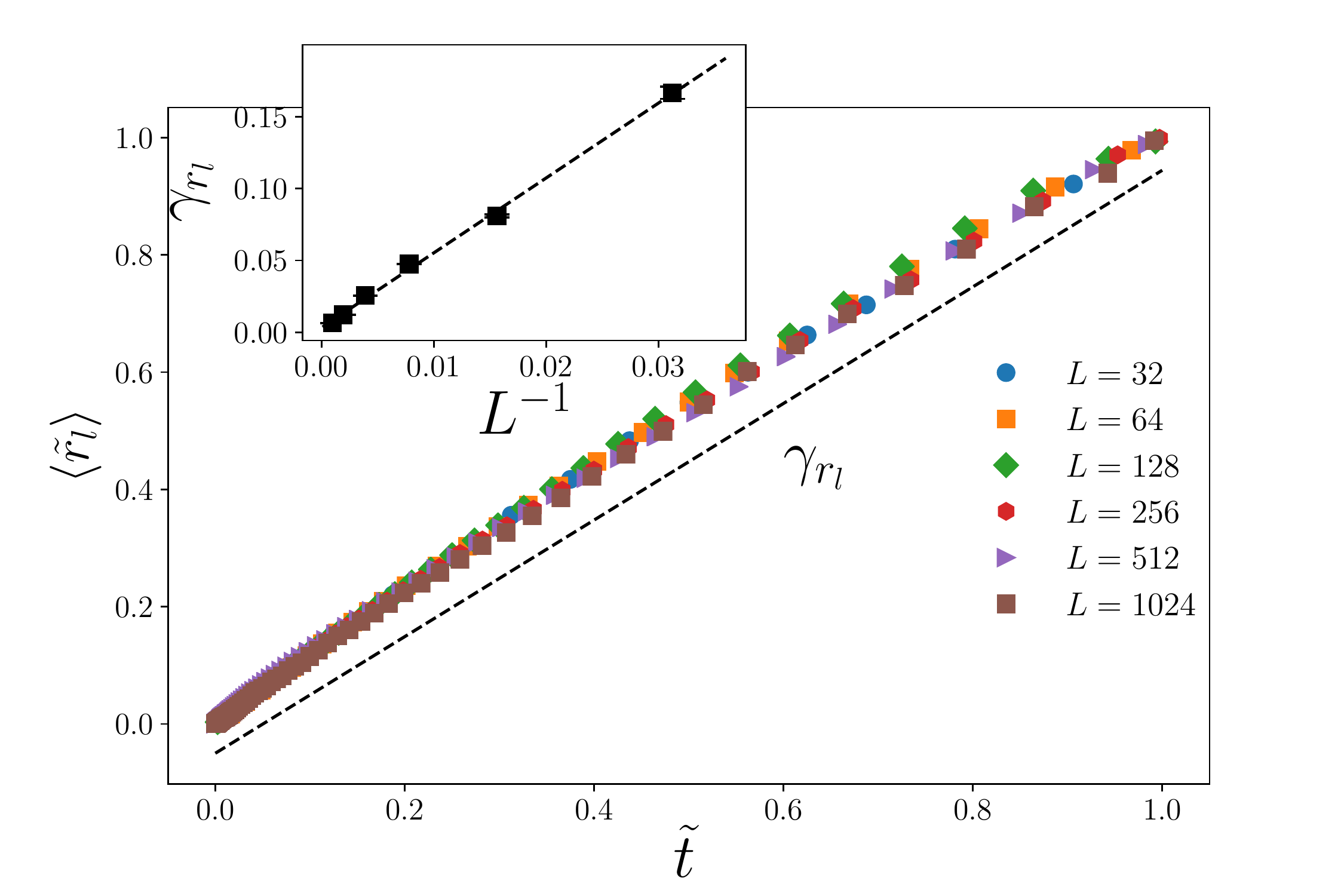}
		\caption{}
		\label{Fig:t_rllarge}
	\end{subfigure}
	\begin{subfigure}{0.45\textwidth}\includegraphics[width=\textwidth]{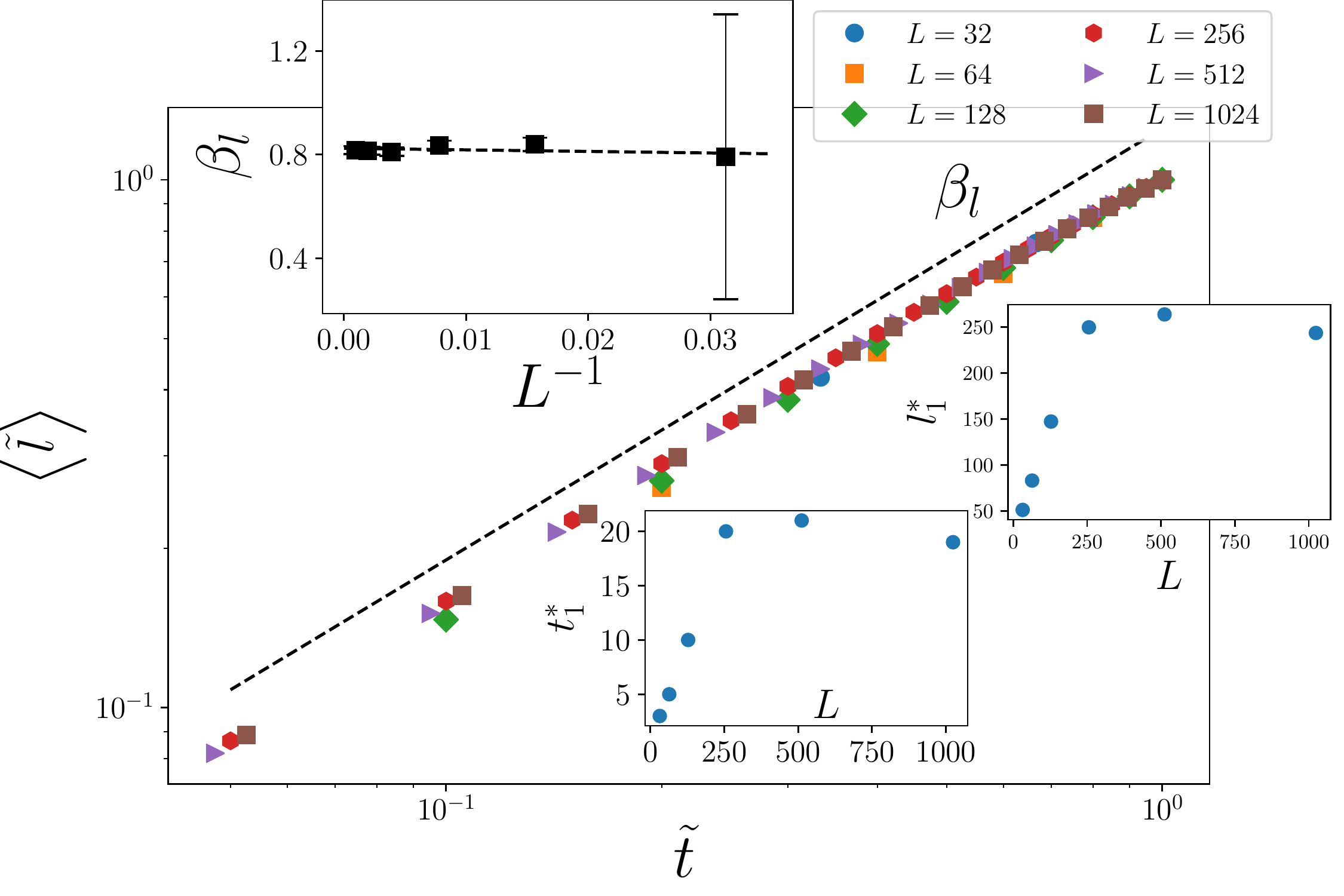}
		\caption{}
		\label{Fig:tlsmallnn}
	\end{subfigure}
	\begin{subfigure}{0.45\textwidth}\includegraphics[width=\textwidth]{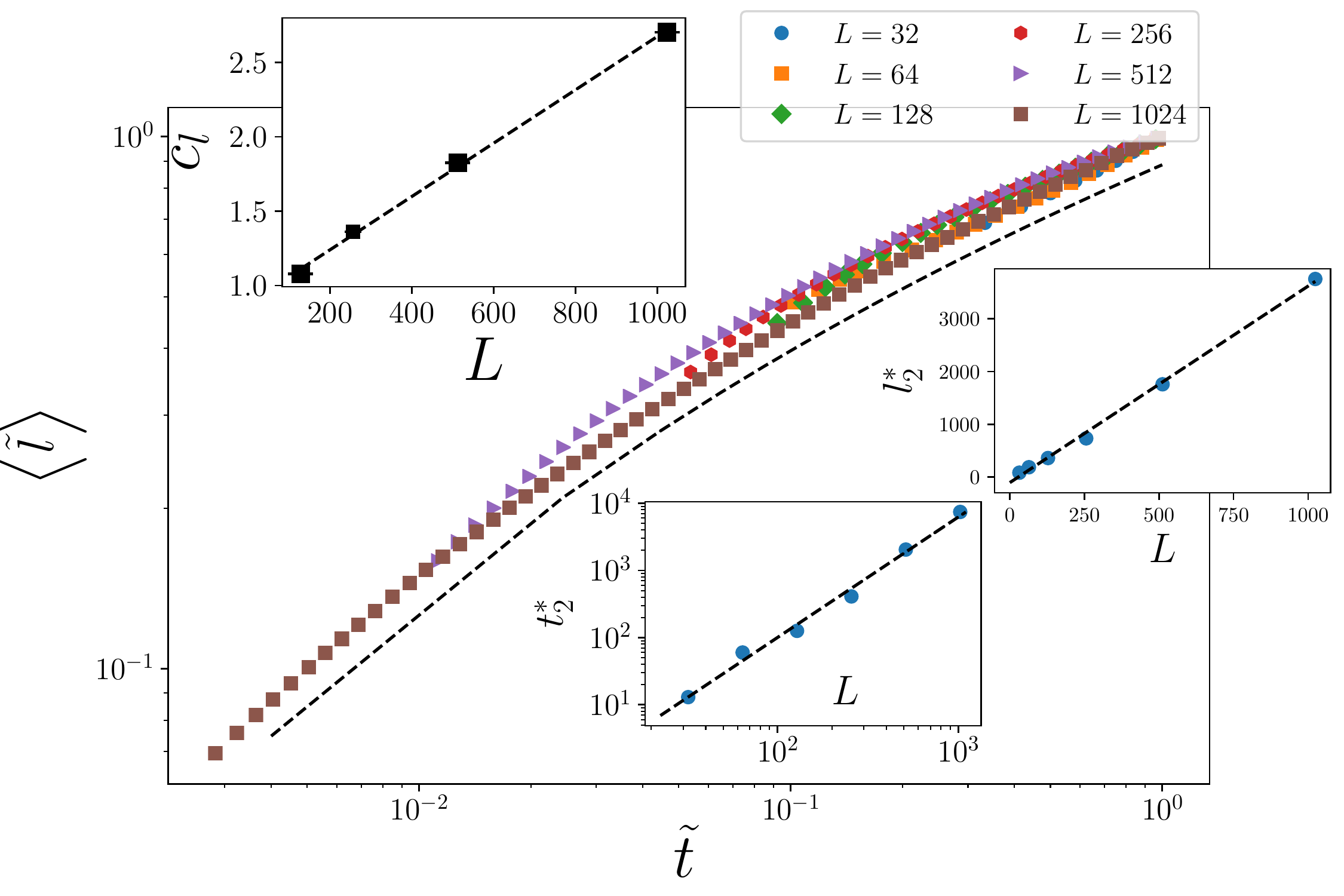}
		\caption{}
		\label{Fig:t_lmiddlenn}
	\end{subfigure}
	\begin{subfigure}{0.45\textwidth}\includegraphics[width=\textwidth]{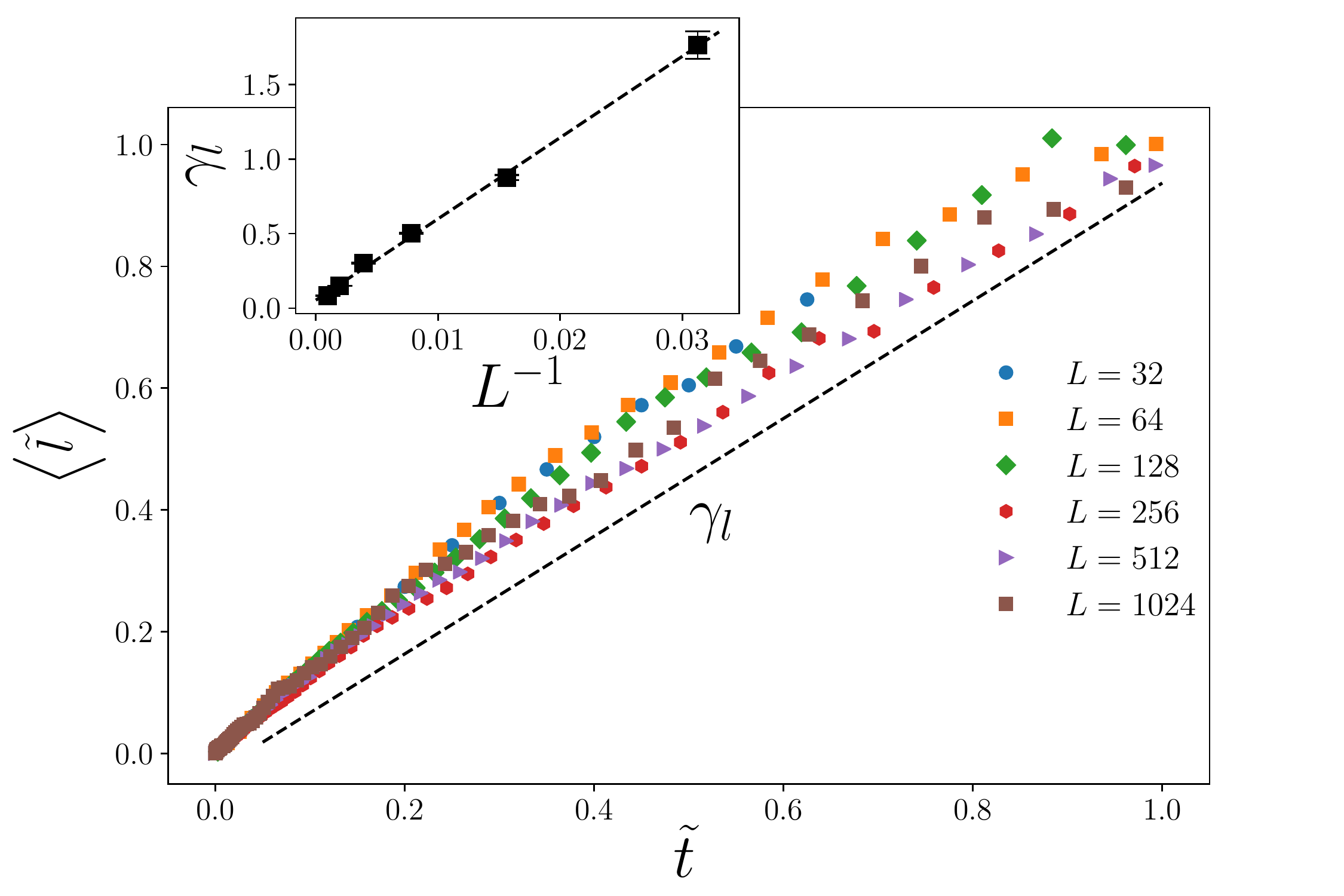}
		\caption{}
		\label{Fig:tllargenns}
	\end{subfigure}
	\caption{(Color Online) The numerical results for random coulomb potential (RCP) background (a)The time dependence of the average of $r_{l}$ in the first phase that renormalize to one. (b)The time dependence of the average of $r_{l}$ in the cross over area. (c)The time dependence of the average of $r_{l}$ in the third phase. (d)The time dependence of the average of $l$ in the first phase that renormalize to one. (e)The time dependence of the average of $l$ in the cross over area. (f)The time dependence of the average of $l$ in the third phase.}
	\label{Fig:trl-tl}
\end{figure*}

Figure~\ref{Fig:t_rlsmallnormalize} shows the gyration radius versus time in the power-law regime, in which it is interestingly seen that $\alpha^{\text{RCP}}_{r_l}(L\rightarrow\infty)=0.73\pm 0.04$, i.e. the system is super-diffussive. This should be compared with ordinary IP which is diffussive (see previous section). We call the second regime as the logarithmic regime since the observables behave like the following relation in this regime
\begin{equation}
	\left\langle x \right\rangle  = a{(\log  (t   + b))^{{c_x}}},
\end{equation}
where $a$ and $b$ are some unimportant constants, and $c_x$ is an exponent. Figure~\ref{Fig:t_rlmiddle} shows the results for $r_l$ in this regime, from which we observe that $c_{r_l}$ becomes $0.5863\pm 0.15675$ in the thermodynamic limit. The lower inset shows that $t^*_2$ goes to infinity faster than $t^*_1$ as $L\rightarrow\infty$, showing that this regime dominates in the thermodynamic limit. The linear regime is described by the linear relation
\begin{equation}
	\left\langle x \right\rangle  = {\gamma _x} t  + b 
\end{equation}
where $\gamma_x$ is the slope, shown in the upper inset of Fig.~\ref{Fig:t_rllarge} which extrapolates to zero in the $L\rightarrow$ limit for $r_l$. In this figure we have plotted re-normalized $\tilde{Y}=\frac{Y}{Y_max}$ versus $\tilde{X}=\frac{X}{X_max}$, where $Y \equiv \frac{{y - {y^ * }}}{\gamma_{r_l}} $, $ X \equiv x - {x^ * }$, and  $({x^ * },{y^ * })$ is crossover point. This analysis shows that the linear regime is actually a stationary regime in the thermodynamic limit, where the quantities are statistically constant. 

The same features are seen for $l$ in Figs.~\ref{Fig:tlsmallnn},~\ref{Fig:t_lmiddlenn}, and~\ref{Fig:tllargenns} from which we see that  $\alpha_l^{\text{RCP}}\rightarrow 0.82\pm 0.03$, and $c_l\rightarrow 0.8815\pm0.1283$ and $\gamma_l\rightarrow 0$ in the limit $L\rightarrow\infty$. The same analysis for $w$ shows the same behaviors, with the exponents $\alpha_w^{\text{RCP}}\rightarrow 0.76\pm 0.05$, and $c_w\rightarrow 0.3058\pm0.45$ and $\gamma_w\rightarrow 0$ in this limit.

\section{conclusion}
In this article, we investigate the effect of short-range and long-range noise on the background of invasion percolation. For this purpose, we have studied the dynamic parameters and the fractal dimension. As observed, the presence of short-range noise in the background of invasion percolation has not shown a significant effect on the behavior of dynamic parameters and fractal dimension, and the behaviors are as strong as standard invasion percolation. The presence of long-range noise in the background of invasion percolation has resulted in the appearance of a crossover area in the dynamic parameters, and the numerical value corresponding to the fractal dimension also shows a significant decrease. This reduction in the fractal dimension indicates the softening of the curves investigated for the largest lattice hole.\\

\appendix
\section{Some of the graphs}\label{SEC:graph}
\begin{figure*}
	\begin{subfigure}{0.47\textwidth}\includegraphics[width=\textwidth]{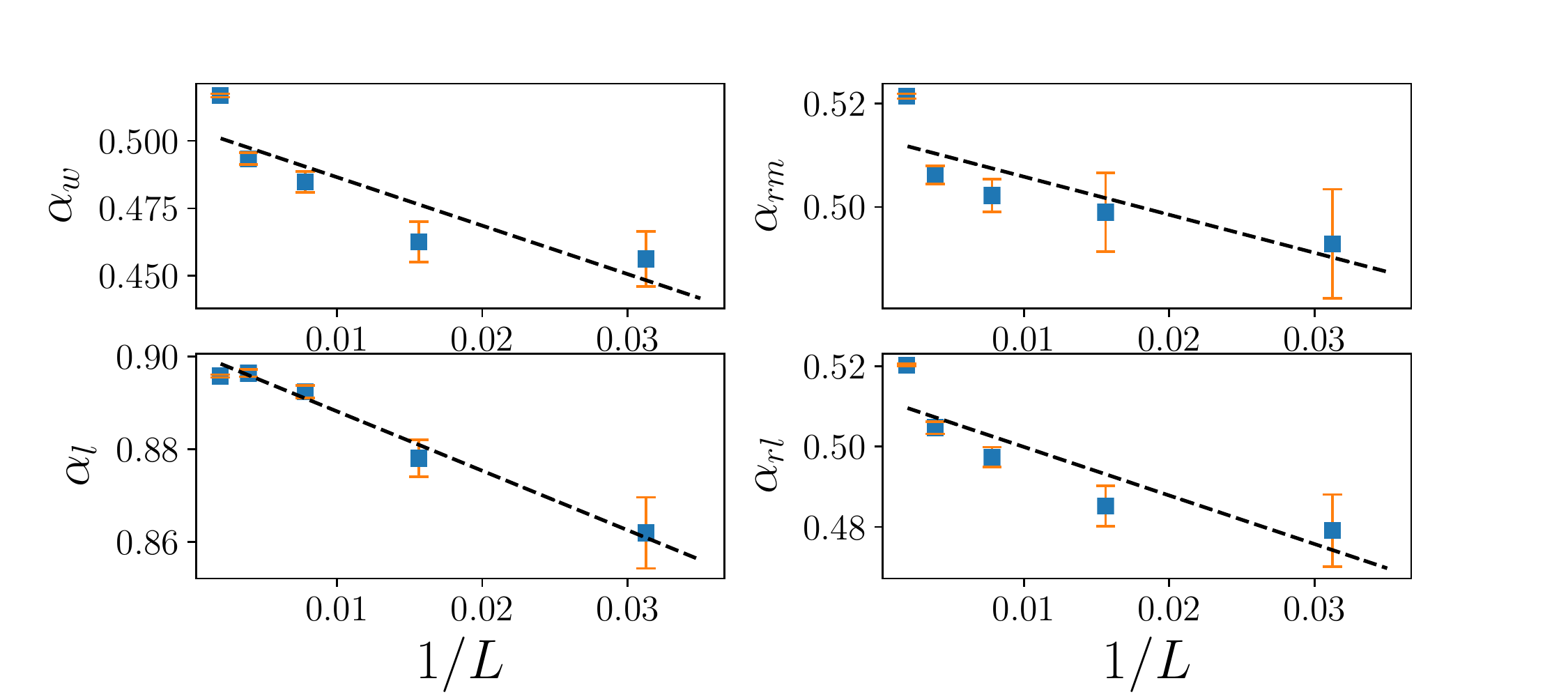}
		\caption{}
		\label{fig:10}
	\end{subfigure}
	\begin{subfigure}{0.47\textwidth}\includegraphics[width=\textwidth]{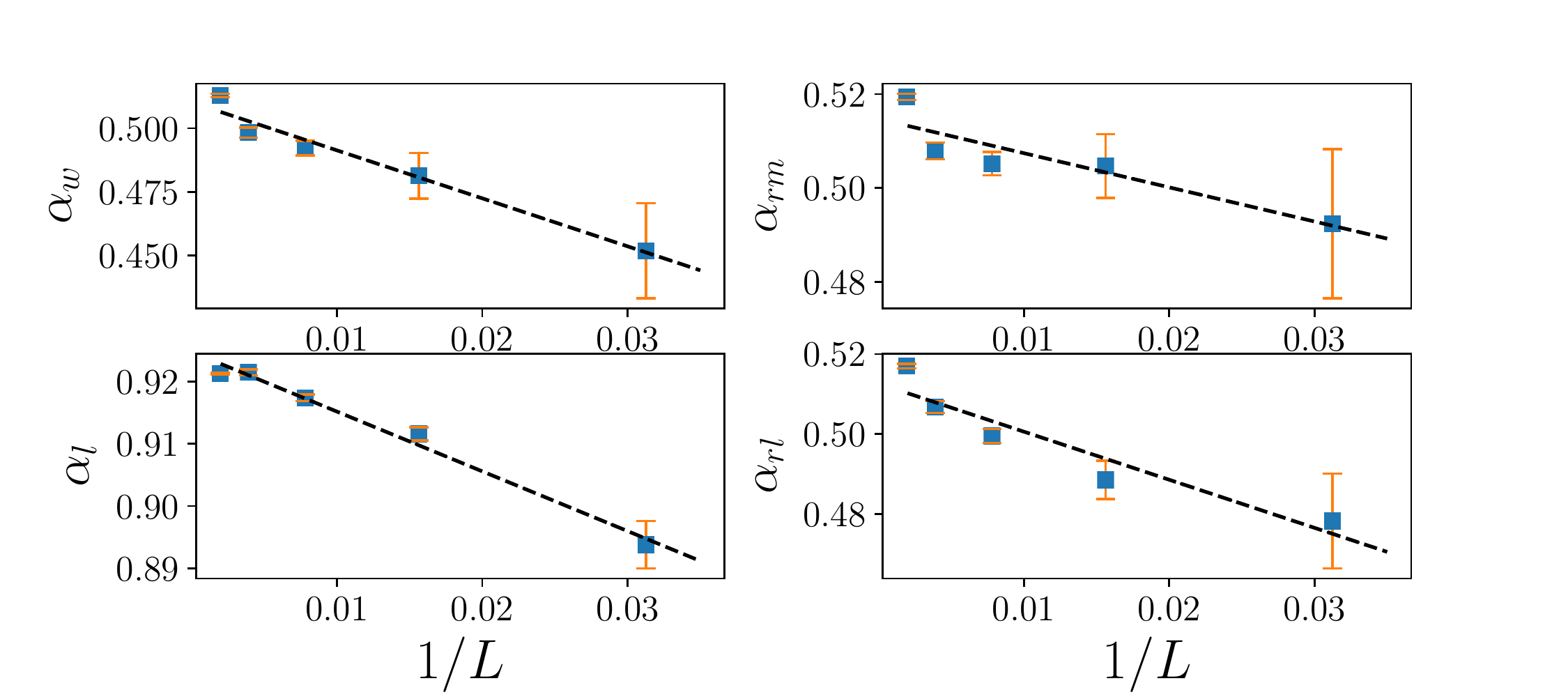}
		\caption{}
		\label{fig:9}
	\end{subfigure}
		\caption{(Color Online) (a) $\alpha_{x}$ in term of $1/L$ for $T=T_{c}$($T_{c}=2.26918$). (b)$\alpha_{x}$ in term of $1/L$ for $p=p_{c}$($p_{c}=0.59275$).(note that $x=rl,rm,l,w$)}
	\label{alphaphix}
\end{figure*}

\begin{figure*}
	\begin{subfigure}{0.32\textwidth}\includegraphics[width=\textwidth]{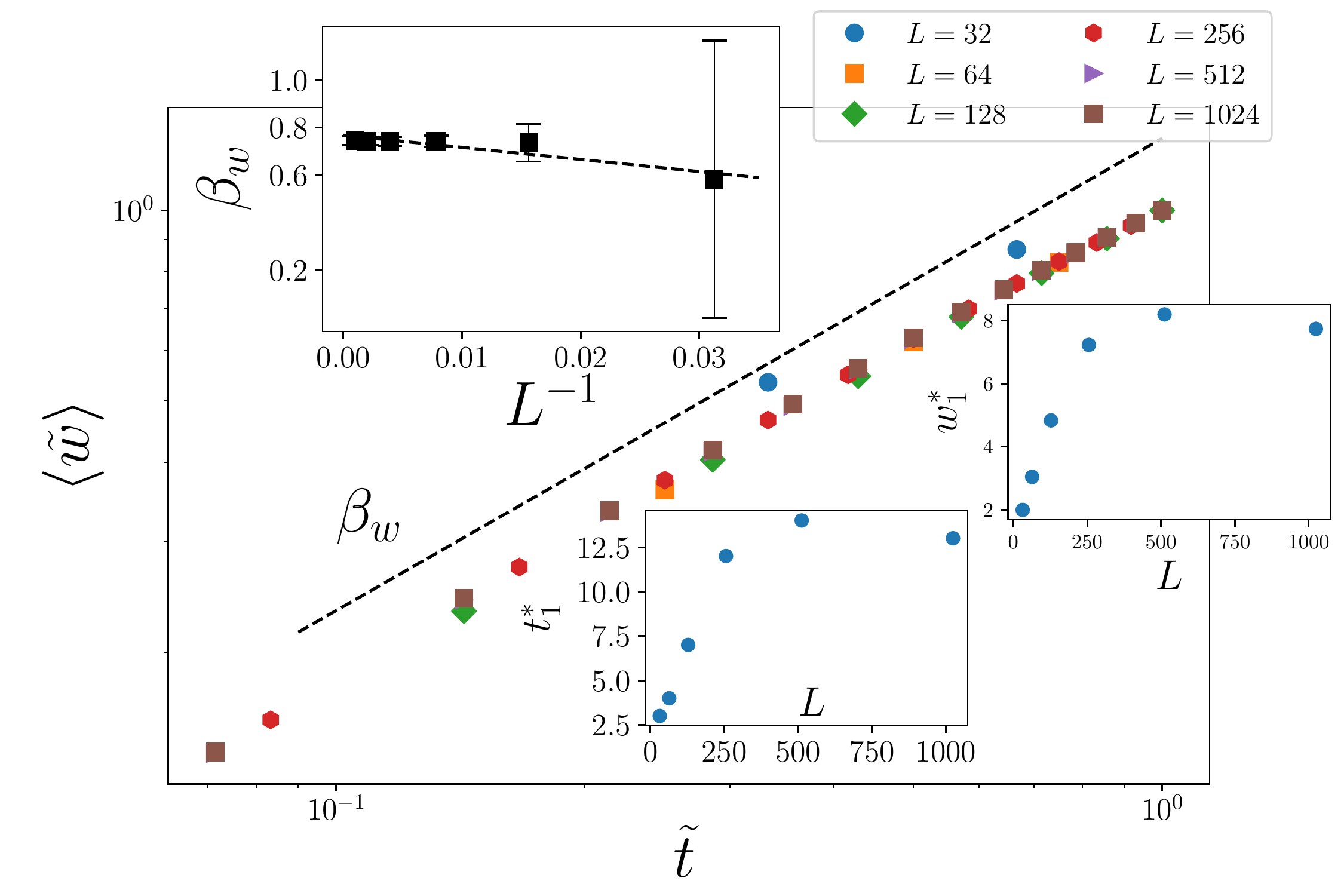}
		\caption{}
		\label{tws}
	\end{subfigure}
	\begin{subfigure}{0.32\textwidth}\includegraphics[width=\textwidth]{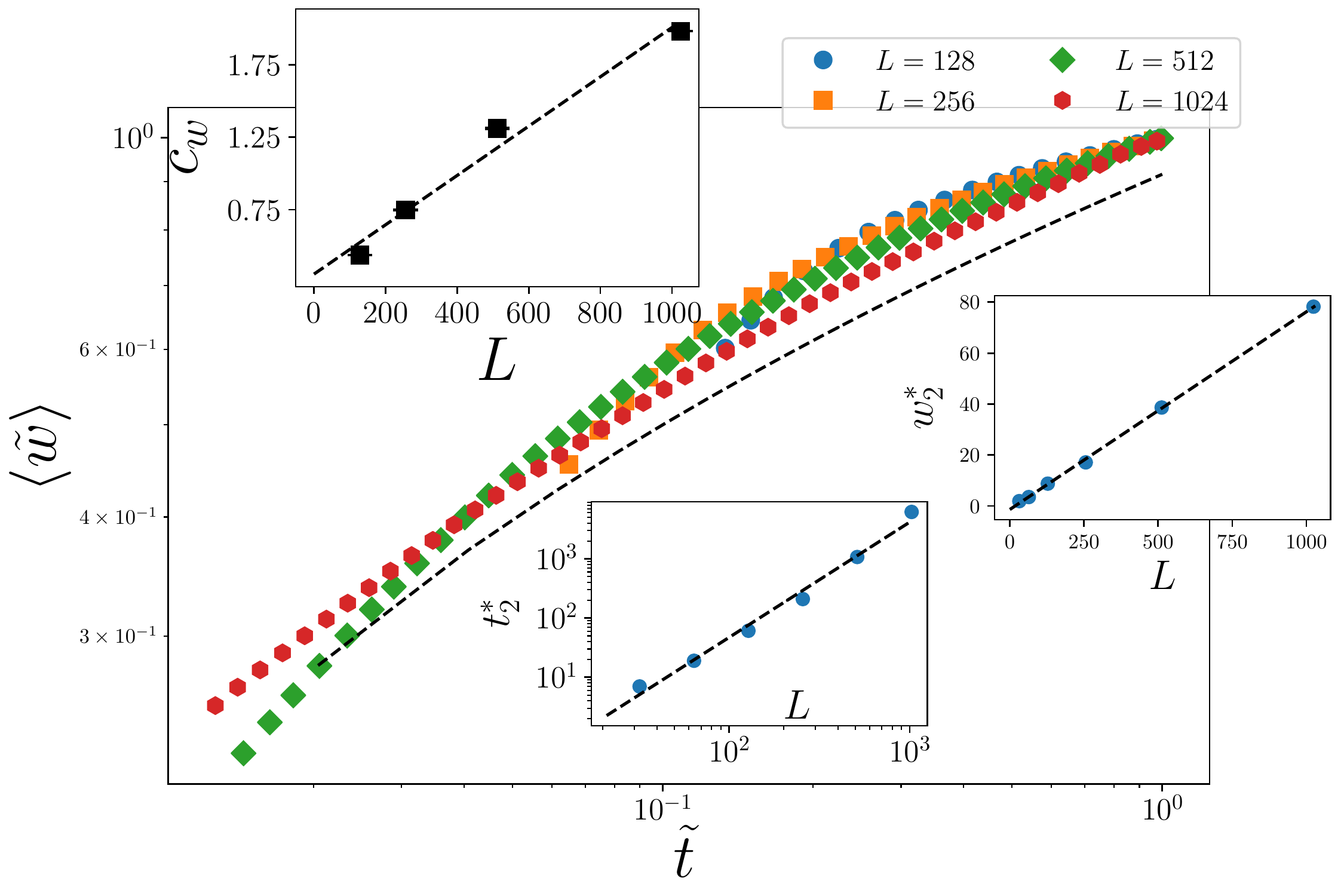}
		\caption{}
		\label{twm2}
	\end{subfigure}
	\begin{subfigure}{0.32\textwidth}\includegraphics[width=\textwidth]{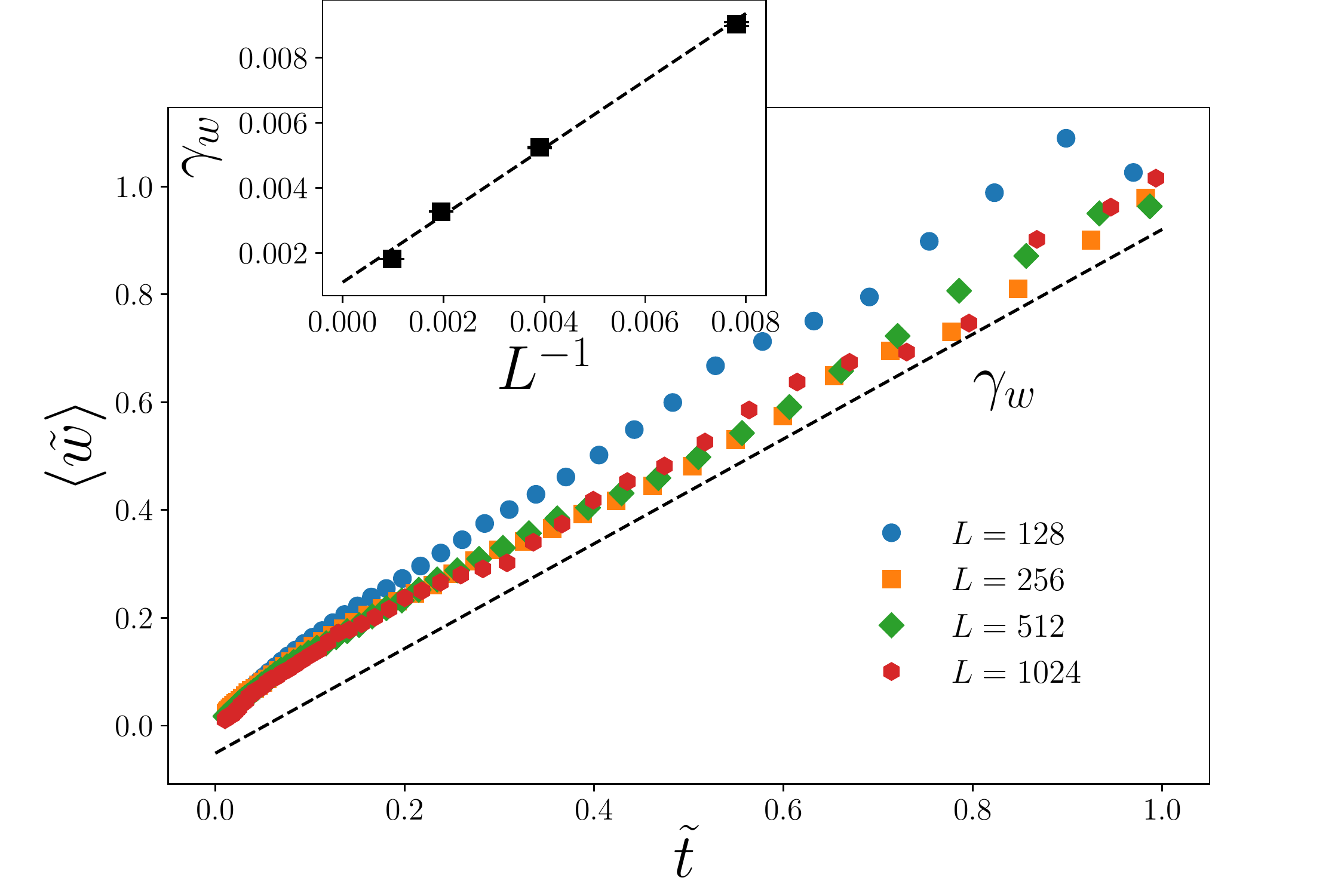}
		\caption{}
		\label{twl}
	\end{subfigure}
	\caption{(Color Online) The numerical results for random coulomb potential (RCP) background (a)The time dependence of the average of $w$ in the first phase that renormalize to one. (b)The time dependence of the average of $w$ in the cross over area. (c)The time dependence of the average of $w$ in the third phase.}
	\label{Fig:tww}
\end{figure*}

\begin{figure*}
	\begin{subfigure}{0.32\textwidth}\includegraphics[width=\textwidth]{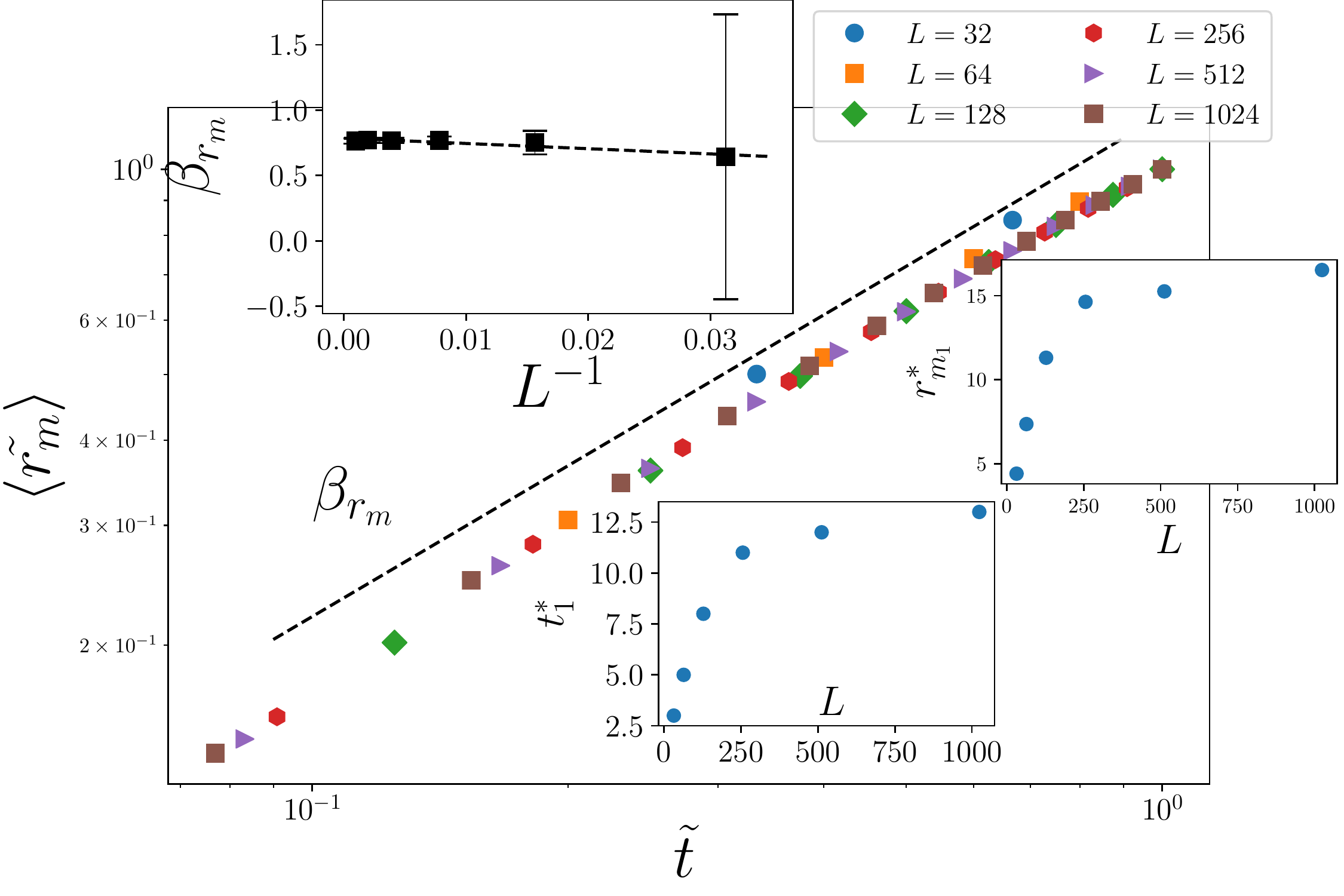}
		\caption{}
		\label{trms}
	\end{subfigure}
	\begin{subfigure}{0.32\textwidth}\includegraphics[width=\textwidth]{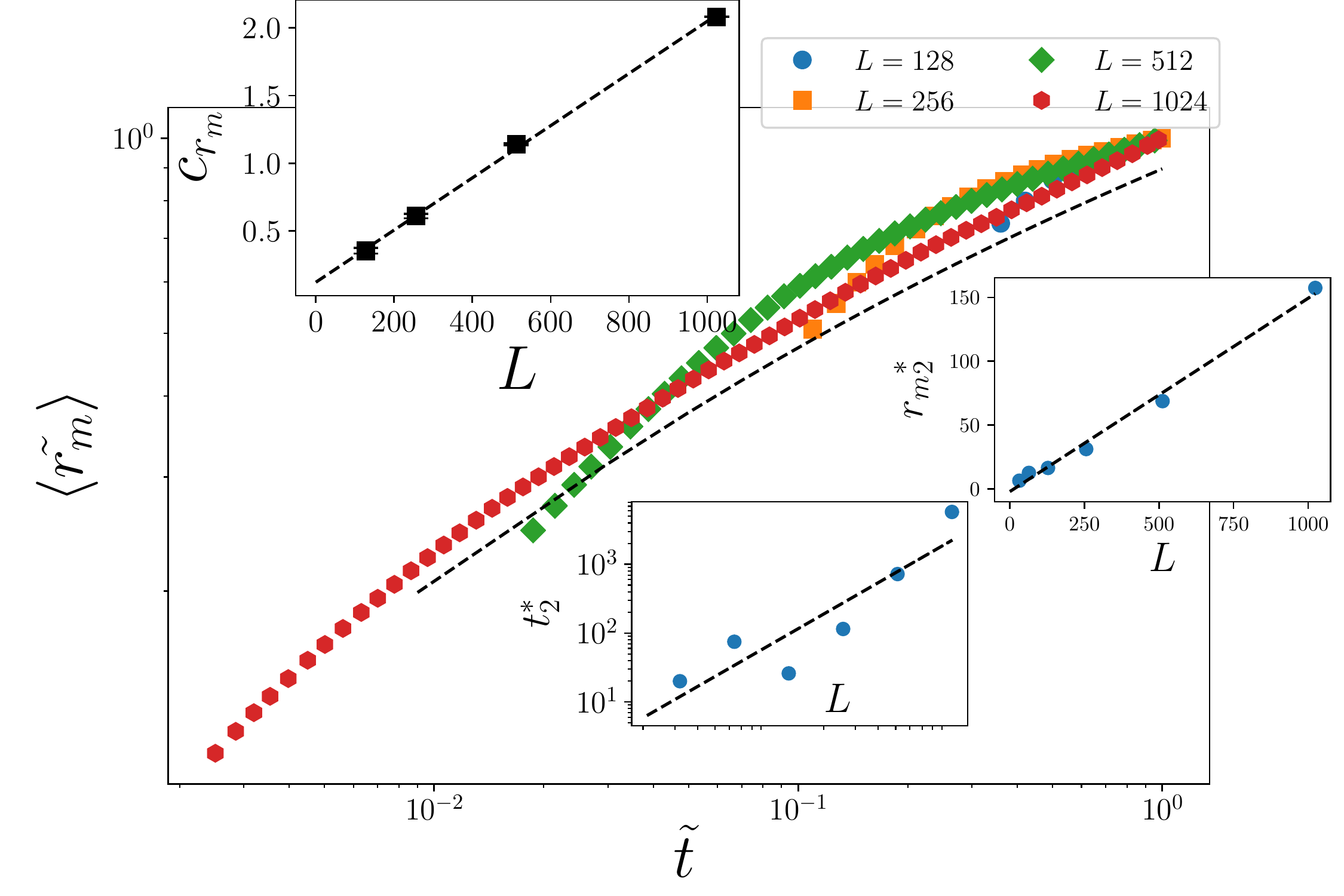}
	\caption{}
	\label{trmm2}
\end{subfigure}
	\begin{subfigure}{0.32\textwidth}\includegraphics[width=\textwidth]{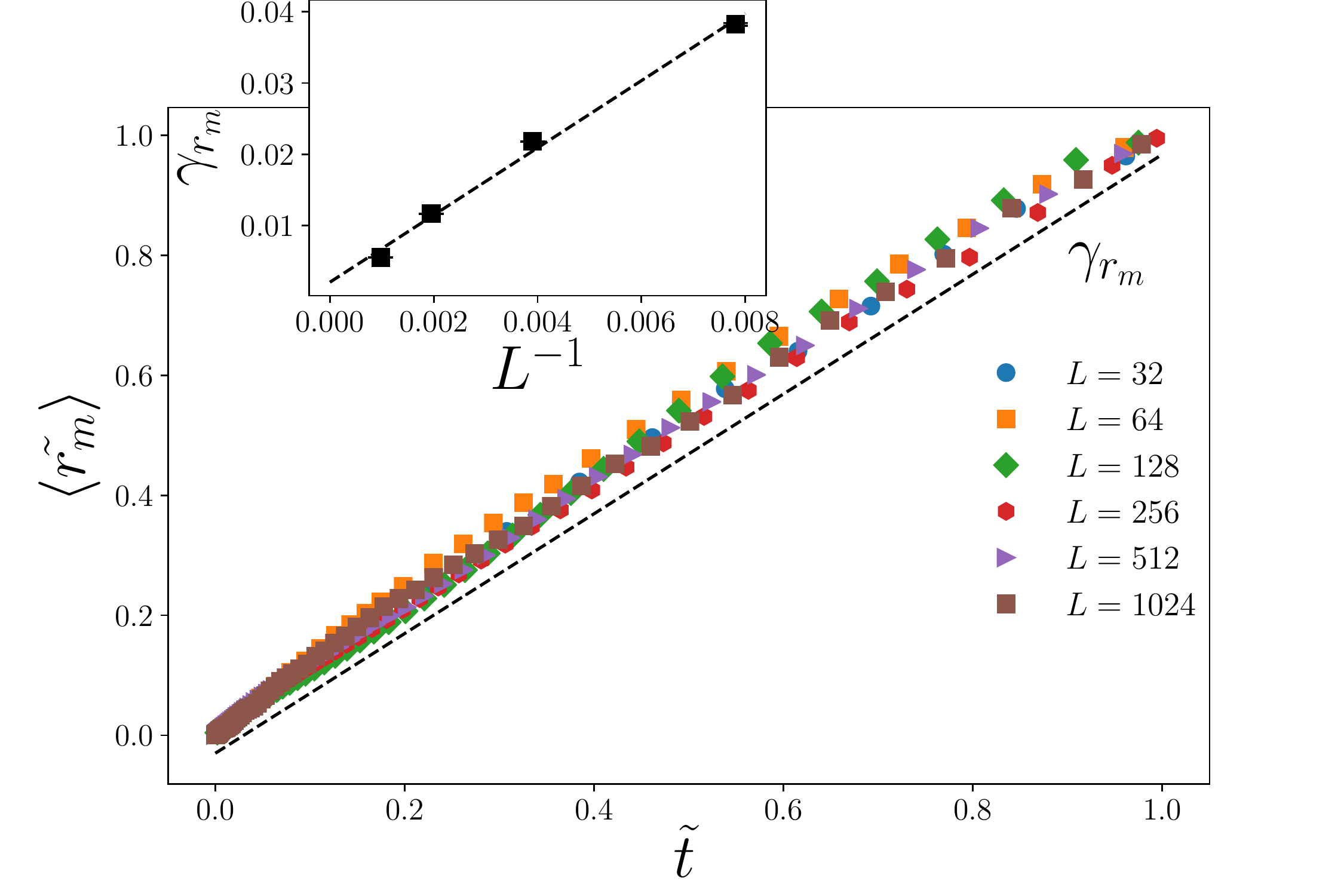}
		\caption{}
		\label{trml}
	\end{subfigure}
	\caption{(Color Online) The numerical results for random coulomb potential (RCP) background (a)The time dependence of the average of $r_{m}$ in the first phase that renormalize to one. (b)The time dependence of the average of $r_{m}$ in the cross over area. (c)The time dependence of the average of $r_{m}$ in the third phase.}
	\label{trm}
\end{figure*}

For more details on short-range notions background, you can refer to Fig.\ref{alphaphix} to see the fit $\alpha_{x}$ in terms of $1/L$ for $T_{c} $ and $p_{c} $. \\
Also, because of the similar behavior of all dynamic processes in the random coulomb potential background, we present graphs of the length and radius of mass gyrus for all three elementary, middle, and final times in this appendix (Fig.\ref{Fig:tww} and Fig.\ref{trm}).


\bibliography{refs}

\end{document}